\documentclass[useAMS]{mn2e}
\usepackage{graphicx}

\title[Chemical compositions of a sample of candidate post-AGB stars]{Chemical composition of a sample of candidate post-AGB stars}
\author[S. Sumangala Rao, Sunetra Giridhar and David L. Lambert]{S. Sumangala Rao$^{1}$\thanks{E-mail:
sumangala@iiap.res.in (SR); giridhar@iiap.res.in (SG); dll@astro.as.utexas.edu (DLL)}, Sunetra Giridhar$^{1}$ and David L. Lambert$^{2}$ \\
$^{1}$Indian Institute of Astrophysics, Bangalore 560034, India\\
$^{2}$ W. J. McDonald Observatory, The University of Texas, Austin, Texas, USA 78712} 
\begin{document}

\pagerange{\pageref{firstpage}--\pageref{lastpage}} \pubyear{2011}

\maketitle

\label{firstpage}

\begin{abstract}

 We have derived elemental abundances for a sample of nine IRAS
     sources with colours similar to those of post-AGB stars. For
     IRAS 01259+6823, IRAS 05208-2035, IRAS 04535+3747 and IRAS 08187-1905 this
     is the first detailed abundance analysis based upon high resolution
     spectra. Mild indication of s-processing for
     IRAS 01259+6823, IRAS 05208-2035 and  IRAS 08187-1905 
     have been found and  a more comprehensive study of s-process
     enhanced objects IRAS 17279-1119 and IRAS 22223+4327 have been  carried out.

     We have also made a contemporary abundance analysis of the high galactic latitude supergiants
       BD+39$^{o}$ 4926  and HD 107369. The former is heavily depleted in refractories
     and estimated [Zn/H] of $-$0.7 dex  most likely gives initial metallicity of the star.
    For HD 107369 the abundances of $\alpha$ and Fe-peak
     elements are similar to  those of halo objects and moderate deficiency of
     s-process elements is seen. IRAS 07140-2321  despite being a
     short period binary with circumstellar shell does not exhibit selective
     depletion of refractory elements.

      We have compiled the  stellar parameters and
     abundances  for post-AGB stars  with  s-process
  enhancement, those showing significant depletion of condensable elements and
   those showing neither.
     The compilation shows that the  s-process enhanced group contains
     very small number of binaries,
     and  observed [$\alpha$/Fe] are generally similar to thick disc values.
     It is likely that they represent AGB evolution of single stars.

     The compilation of depleted group contains larger fraction of binaries and
      generally supports the hypothesis of dusty discs surrounding binary post-AGB stars
      inferred via the shape of their SED and mid IR interferometry.
      IRAS 07140-2321 and BD+39$^{o}$ 4926  are difficult to explain with
      this scenario and indicate the existence of additional
      parameter/condition needed to explain the depletion phenomenon.
     However  the conditions for  discernible depletion, minimum temperature of 
    5000K and initial metallicity larger than $-$1.0 dex 
      found from our earlier work still serves as useful criteria.

 {\it Subject headings: stars:abundances -- stars:AGB and post-AGB --
stars: variables:other (RV\,Tauri)}

\end{abstract}


\section{Introduction}

Considerable theoretical and observational interest has in
recent years been focused on asymptotic giant branch (AGB)
stars but much remains mysterious about them. The AGB phase is
enjoyed by low and intermediate
 mass stars (approximate mass range 0.8 to 8M$_{\odot}$). It is in this phase
that a star experiences extensive internal nucleosynthesis whose
fruits are dredged to the stellar surface {\bf (see Herwig 2005)}. Furthermore, mass-loss
ensures that the products of nucleosynthesis are dispersed into the
circumstellar and subsequently the interstellar environment. Thus, AGB
stars are likely major contributors of Li, C, N, F and $s$-process
elements among others to Galactic chemical evolution {\bf (Romano et al. 2010)}.

Observational validation of theoretical investigations of how
AGB stars achieve internal nucleosynthesis, dredge-up and
mass-loss are generally hampered by the fact that the more evolved
and more interesting of these stars have low temperature
atmospheres and therefore spectra replete with dense and
complex molecular {\bf lines}. Although these spectra have been
analysed by a few stellar spectroscopists, the abundance studies
are generally limited to very few elements and isotopic ratios.

The AGB phase of stellar evolution concludes with a rapid phase
of evolution to the tip of the white dwarf cooling track. This
post-AGB (PAGB) phase takes a star in a few thousand years or so
from a cool AGB star to a very hot central star of a planetary
nebula and onto a white dwarf cooling track. Along this track, the
star's spectrum is amenable to straightforward abundance analysis
and thus offers apparently a way to infer the abundance changes
brought by the PAGB star's AGB progenitor. This inference is, of course,
dependent on the assumption that the composition of an AGB star is
exactly preserved by the PAGB star. Observations of certain PAGB stars
show that this is a false assumption. For example, several PAGB
stars show abundance anomalies correlated with the condensation
temperature (T$_C$)\footnote{\bf The condensation temperature T$_C$ is the temperature at which half of a particular element in a gaseous environment condenses into dust grains. Condensable elements refer to the refractories like Ti, Ca, Sc and the s-process elements which readily condense into dust grains because of their high T$_C$.} for dust grains to condense out of gas of normal
composition (see Van Winckel 2003 for a review)
 -- a process we refer to as `dust-gas winnowing'.
Other PAGB candidate stars show abundance
anomalies correlated with the ionization potential of the neutral
atoms (Rao \& Reddy 2005) -- a process we refer to as `the FIP effect'.

Given that the PAGB phase is rapid, the number of identified PAGB
stars is relatively small and the number subjected to an
abundance analysis is, of course, even smaller. In this paper,
we report abundance analyses for 11 candidate PAGB stars and
have compiled abundance data of previously analysed PAGB stars in an
attempt to seek explanations for the diverse compositions of these
objects.

\begin{figure}
\begin{center}
\includegraphics[width=10cm,height=10cm]{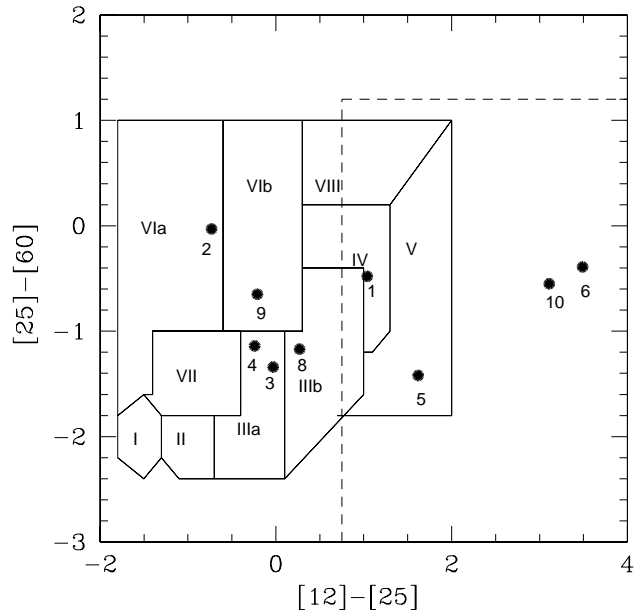}
\caption{The location of sample stars in IRAS color-color diagram
containing the zones defined by van der Veen $\&$ Habing (1988) and area enclosed in the
 dashed line which according to Szczerba et al. 2007 is richly populated by PAGB stars with
 colours like planetary nebulae.
 The sample stars are numbered as follows:
1- IRAS 01259+6823
2- IRAS 04535+3747
3- IRAS 05208-2035
4- IRAS 07140-2321
5- IRAS 07331+0021
6- IRAS 08187-1905
8- IRAS 12538-2611
 9- IRAS 17279-1119
 10- IRAS 22223+4327
}
  \label{loci1}
\end{center}
\end{figure}
 \section{Selection of the  sample}

Our sample stars are  presented in Table 1 and displayed in the
IRAS two colour diagram Figure \ref {loci1} which has proven to be
a powerful tool for identifying candidate PAGB stars (van
der Veen \& Habing 1988;
Su\'{a}rez et al. 2006; Szczerba et al. 2007). Nine of our eleven stars
have measured IRAS fluxes and are identified in Figure \ref{loci1}; the
stars, HD 107369 and BD +39$^{o}$ 4926 were not detected by IRAS.

It is known that zone 1 of IRAS two colour diagram by van
der Veen \& Habing 1988 corresponds to fluxes from stellar photospheres
warmer than 2000K. The IRAS colours of Zones II-III signify the
emergence and evolution of circumstellar shell (CS) produced by
increasing large mass-loss at AGB. Sources at Zone IV are at the
super-wind phases or slightly beyond. Zone V contains objects with only
signature of cold dust shell as the mass-loss has stopped hence warm dust
is not being added. The objects with detached cold CS would be seen at region
VIa, while objects in zone VIb contains objects with warm as well as cold
CS. The Zone VII and dashed region is dominated by PAGBs
with PN like colours.

Our sample stars IRAS 05208-2035,
IRAS 07140-2321, IRAS 12538-2611 and IRAS 01259+6823
 belong to zone IIIa, IIIb and IV
 where signatures of CS formed by increased mass-loss at late AGB are evident.
 IRAS 17279-1119 (zone VIb) seems to possess the hot as well as cold dust
shell while IRAS 07331+0021, IRAS 22223+4327, IRAS 08187-1905
 have PN like colours.

\begin{table*}
 \centering
 \begin{minipage}{140mm}
  \caption{The Program Stars.}
 \label{table1}
\begin{tabular}{lllll}
  \hline
\multicolumn{1}{l}{No.}&  
\multicolumn{1}{l}{IRAS}&  
\multicolumn{1}{l}{Other Names}&
\multicolumn{1}{l}{Var Type}&
\multicolumn{1}{l}{Colors}\\

 \hline
1& 01259+6823   &  ... &...& PN like   \\
2& 04535+3747&  V409 Aur, HD 280138, BD+37$^{o}$996& SRD& ...  \\
3& 05208-2035&  BD-20$^{o}$1073& ...& RV Tauri like      \\
4& 07140-2321&  CD-23$^{o}$5180, SAO 173329& Irregular& RV Tauri like    \\
5& 07331+0021 & AI CMi, BD+00$^{o}$ 2006& Irregular& ...     \\
6& 08187-1905 & V552 Pup, HD 70379, BD-18$^{o}$ 2290 & SRD& ...  \\
7& ...  & HD107369, SAO 203367& ... & ...   \\
8& 12538-2611 & HR 4912, HD 112374, BD-18$^{o}$ 2290& SRD& RV Tauri like  \\
9& 17279-1119 &HD 158616, BD-11$^{o}$4391, V340 Ser& RV Tauri& RV Tauri like \\
10& 22223+4327 & V448 Lac, BD +42$^{o}$ 4388& SRD& PN like  \\
11&       ...  & BD+39$^{o}$4926, SAO 72704& ..& ... \\
\hline
\end{tabular}
\end{minipage}
\end{table*}

\begin{table*}
 \centering
 \caption{Stellar Parameters Derived from the
Fe-line Analyses}
\label{table2}
\begin{tabular}{lllcrlcrlr}
 \hline
\multicolumn{1}{l}{Star} &
\multicolumn{1}{l}{UT Date}&
\multicolumn{1}{l}{V$_{r}^{\rm a}$}&
\multicolumn{1}{c}{T$_{\rm eff}$, $\log g$,
[Fe/H]} &
 \multicolumn{1}{c}{$\xi^{\rm b}_{\rm t}$} & 
 \multicolumn{2}{c}{Fe
I$^{\rm c}$} & & 
\multicolumn{2}{c}{Fe II$^{\rm c}$}   \\ \cline{6-7} \cline{9-10}
&& \multicolumn{1}{l}{(km~s$^{-1}$)} &&
\multicolumn{1}{l}{(km~s$^{-1}$)} &
\multicolumn{1}{l}{$\log \epsilon$}& \multicolumn{1}{l}{n} & & \multicolumn{1}{l}{$\log \epsilon$} &\multicolumn{1}{l}{n} \\ 
 \hline
IRAS01259+6823&2007 Nov 5& $-$49.1& 5000,1.50,$-$0.60 & 3.3\phantom{000}& $6.82\pm 0.10$ &
60 && $6.88 \pm 0.11$ &  11 \\
IRAS04535+3747&2008 Feb29& $-$37.4& 6000,1.25,$-$0.48 & 3.6\phantom{000}& $6.96\pm 0.19$ &
97 && $6.99 \pm 0.10$ &  12 \\
IRAS05208-2035 & 2007 Nov 3 & $+$50.3& 4250, 0.75, $-$0.65 & 1.6\phantom{000} & $6.80 \pm 0.20$ &
  49& &$6.81 \pm 0.11$ & 7 \\
IRAS07140-2321 & 2007 Nov 2& $+$68.0& 7000,1.00, $-$0.92 & 3.6\phantom{000}&$6.55 \pm 0.14$ &
75 && $6.52 \pm 0.11$  &19\\
IRAS07331+0021 & 2008 Apr 19 & $+$47.9& 4500, 1.00 ,$-$1.16 & 5.2\phantom{000} & $6.27 \pm 0.16$ &
26 && $6.32 \pm 0.10$ & 4 \\
IRAS08187-1905 & 2008 Feb 26& $+$ 61.5& 6250,0.50, $-$0.59 & 3.8\phantom{000} & $6.90 \pm 0.15$ &
33 && $6.82 \pm 0.13$  &11 \\
HD107369 & 2008 Apr 20 &$-$39.0& 7500, 1.50, $-$1.33 & 1.3\phantom{000} & $6.09 \pm 0.13$ &
19 && $6.15 \pm 0.12$  &30 \\
IRAS12538-2611 & 2009 May 10& $-$31.3& 5250,1.00, $-$1.12 & 4.6\phantom{000} & $6.34 \pm 0.11$ &
59 && $6.33 \pm 0.11$  &12  \\
IRAS17279-1119 &2008 Aug 10& $+$68.7& 7250, 2.25, $-$0.43 & 4.7\phantom{000} & $7.02 \pm 0.21$ &
21 & &$7.01 \pm 0.16$ & 12 \\
IRAS22223+4327 & 2009 Dec 27& $-$41.3& 6500, 1.00, $-$0.33 & 4.3\phantom{000} & $7.14 \pm 0.08$ &
20 && $7.11 \pm 0.10$  &16 \\
BD +39$^{o}$ 4926 &2007 Nov 19& $-$36.3&  7750, 1.00, $-$2.37 & 3.0\phantom{000} & $5.02 \pm
0.15$ & 9 && $5.15 \pm 0.13$ & 21 \\
\hline
\end{tabular}
\flushleft$^{a}${V$_{r}$ is the radial velocity in {km~s$^{-1}$}}
\flushleft$^{b}${$\xi_{\rm t}$ is the microturbulence}
\flushleft$^{c}${$\log \epsilon$ is the mean abundance relative to H
(with $\log \epsilon_{\rm H} = 12.00$).
The standard deviations of the means as calculated
from the line-to-line scatter are given.
$n$ is the number of
considered lines.}
\end{table*}
\section {Observations}

\begin{figure}
\begin{center}
\includegraphics[width=9cm,height=9cm]{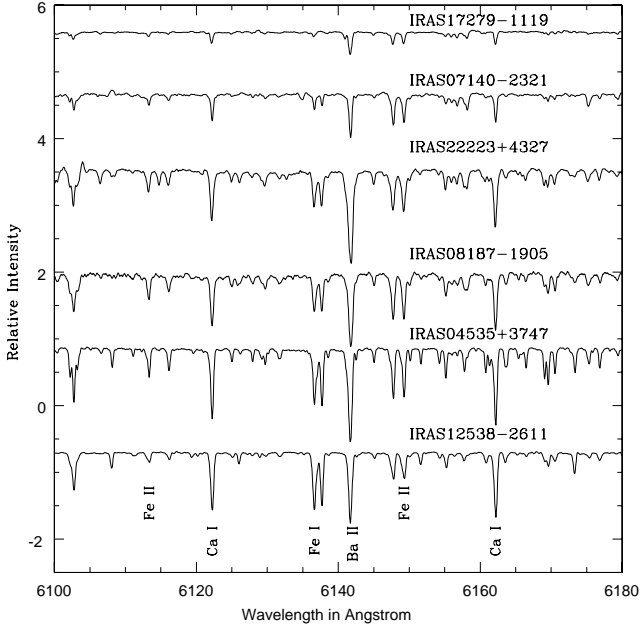}
\caption{Sample spectra of our stars presented in descending order of
 temperature (top to bottom) in the 6100-6180 region.}
\label{loci2}
\end{center}
\end{figure}

High-resolution optical spectra were obtained at the W.J. McDonald
Observatory with the 2.7m Harlan J.
 Smith reflector and the Tull coud\'e spectrograph (Tull et al. 1995)
with a resolving power of 60,000.
Spectral coverage  in a single exposure
from this cross-dispersed echelle spectrograph
is complete up to 6000\AA~ and extensive but incomplete at
longer wavelengths.
A S/N ratio of 80-100 over much of the spectral range was achieved.
 Figure \ref {loci2} contains a few representative spectra to illustrate
 the quality of typical spectra.

 \section{Abundance analysis}
 
     We have made use of the new grid of ATLAS09 model atmospheres available at database
    of Kurucz\footnote{http://kurucz.harvard.edu/grids.html}.
   Spectrum synthesis code MOOG (2009 version) by C. Sneden (1973) has been used.
    The assumptions are standard, Local Thermodynamical Equilibrium (LTE),
   plane parallel atmosphere and  hydrostatic equilibrium and flux conservation.

    The hydrogen lines for most stars were affected by emission components
    hence could not be used for parametrization of stars. We have used
     Fe\,{\sc i} and Fe\,{\sc ii} lines for deriving atmospheric parameters.  Further,
     lines of Mg\,{\sc i}, Mg\,{\sc ii}; Si\,{\sc i}, Si\,{\sc ii} ; Ti\,{\sc i} , Ti\,{\sc ii} and Cr\,{\sc i}, Cr\,{\sc ii} 
    were also used as additional constraints whenever possible.

    Our sample stars were generally hotter than 5000K with the exception of
    IRAS 07331+0021 and IRAS 05208-2035, which fortunately
    turned out to be metal-poor by
   $-$1.2 dex and $-$0.65 dex. Hence adequate number of unsaturated clean
   lines could be measured for number of elements for them.

 First, the microturbulence $\xi_{t}$ is derived by requiring that
  the derived abundances are independent of line strengths.
  We have generally used Fe\,{\sc ii} lines for measuring $\xi_{t}$
  microturbulence for most of our sample stars,
  since appreciable departure from 
  LTE are known to occur for Fe\,{\sc i} lines (Boyarchuk et al.
  1985, Th\'{e}venin \& Idiart 1999).  However for coolest members,
   IRAS 05208-2035 and IRAS 07331+0021 very few Fe\,{\sc ii} lines
   could be measured and hence Fe\,{\sc i} lines had to be used
   instead for fixing the microturbulence.

   The temperature is estimated  by requiring that derived abundances are
   independent of the Lower Excitation Potential (LEP).
   IRAS 07140-2321, HD 107369 and BD+39$^{o}$4926 had large number of useable
  Fe\,{\sc ii} lines. In fact, for  HD 107369 Fe\,{\sc ii} lines cover good LEP
  range to estimate the temperature. For IRAS 07140-2321 and BD+39$^{o}$4926
  the LEP covered by Fe\,{\sc ii} lines was lesser than those by Fe\,{\sc i} hence
  both species were employed although we did not find any difference between
  the temperatures estimated from them. For cool stars the
  observed Fe\,{\sc ii} did not have a good range in LEP hence Fe\,{\sc i} lines were employed.
  The gravity was derived by requiring Fe\,{\sc i} and Fe\,{\sc ii} giving the
  same abundance. In addition, the ionization equilibrium of Mg\,{\sc i}/ Mg\,{\sc ii} ,
  Si\,{\sc i}/Si\,{\sc ii} , Sc\,{\sc i}/Sc\,{\sc ii},
  Ti\,{\sc i}/Ti\,{\sc ii}, Cr\,{\sc i}/Cr\,{\sc ii} were used as additional constraints whenever possible.

The accuracy of equivalent width  measurements depends on resolution, spectral
type of the star and the continuum fitting. For stars of spectral type late A- F,
the equivalent widths could be measured with an accuracy of $\sim$ 5 -8 \% in absence of
line asymmetries.

   With the above mentioned accuracy in measured equivalent widths,
  the microturbulence velocity could be measured with an accuracy
  of $\pm$0.25 km s$^{-1}$, temperature with $\pm$150K and {\itshape log g} of
  $\pm$0.25 cm s$^{-2}$.

But for cooler stars, line strengths could only be measured with
 accuracy of 8-10 \%. The microturbulence velocity could be measured with
an accuracy of $\pm$0.5 km s$^{-1}$, temperature with $\pm$250K and {\itshape log g} of
  $\pm$0.5 cm s$^{-2}$.

The derived atmospheric parameters and  heliocentric radial velocities
 for the epoch of observations
for the program stars are presented in Table 2.

The sensitivity of the derived abundances to the
uncertainties of atmospheric parameters {\itshape T$_{\rm eff}$, log g}, and $\xi$
are summarized in Table 3. For four stars representing
the full temperature range of our sample, we present changes in
[X/Fe] caused by varying atmospheric parameters
by 200K, 0.25 cm s$^{-2}$ and 0.5 km s$^{-1}$ (average
accuracies of these parameters) with respect to the chosen model
for each star.

  The total stellar parameter related error is estimated by taking the
 square root of the sum of the square of the systematic errors (individual errors
 associated with uncertainties in temperatures and gravities) and
bars in the abundance plot correspond to the systematic error.

Having determined the atmospheric parameters, the abundances of
different elements were derived using the available lines.
The derived abundances relative to the solar abundances are presented
in their respective tables. The solar photospheric abundances
given by Asplund, Grevesse \& Sauval (2005) have been used as reference
values.

The sources of the {\itshape gf} values for different elements used in our abundance
analysis have been listed in Table 4.

 We have investigated possible systematic effects caused
 by the adopted {\itshape gf} values from different sources as follows. We have measured solar equivalent widths
 for our lines on  Solar Flux Atlas by  Kurucz et al. (1984) and estimated the
 abundances using model atmosphere appropriate for the sun {\itshape T$_{\rm eff}$}=5770, {\itshape log g} of
 4.4 and $\xi_{t}$ of 0.9 kms$^{-1}$.
 The agreement for most elements is within 0.04 to 0.09 dex although the linelist
 is different for different stars .

The lines of certain elements are affected by hyperfine splitting.
We have  included the hyperfine structure components in our line list while
synthesizing the spectral feature of these elements for
deriving the abundances. For elements Sc and Mn, we have used
  hfs component list and their {\itshape log gf} given by
 Prochaska $\&$ McWilliam (2000), for Cu (Allen $\&$ Porto de Mello 2011),
 for Eu (Mucciarelli et al. 2008) and for
 Ba (McWilliam 1998). The effect of isotopic components
 were considered for Eu II (2 isotopes) and Ba II (5 isotopes) lines.

\begin{table*}
 \centering
 \caption{Sensitivity of [X/Fe] to the uncertainties 
in the model parameters for a range of temperatures covering our sample stars .}
\label{table3}
\begin{tabular}{llcrlcrlcrlcr}
  \hline
&\multicolumn{3}{c}{IRAS 05208-2035}&
\multicolumn{3}{c}{IRAS 12538-2611}&
\multicolumn{3}{c}{IRAS 22223+4327}&
\multicolumn{3}{c}{BD $+$39$^o$4926} \\
\cline{2-13}
&& \multicolumn{1}{l}{(4250K)}
&&& \multicolumn{1}{l}{(5250K)}
&&& \multicolumn{1}{l}{(6500K)}
&&& \multicolumn{1}{l}{(7750K)} \\
\multicolumn{1}{l}{Species} &
\multicolumn{1}{c}{$\Delta$$T_{\rm eff}$}&
\multicolumn{1}{c}{$\Delta$log~$g$}&
\multicolumn{1}{c}{$\Delta \xi$}&
\multicolumn{1}{c}{$\Delta$$T_{\rm eff}$}&
\multicolumn{1}{c}{$\Delta$log~$g$}&
\multicolumn{1}{c}{$\Delta \xi$}&
\multicolumn{1}{c}{$\Delta$$T_{\rm eff}$}&
\multicolumn{1}{c}{$\Delta$log~$g$}&
\multicolumn{1}{c}{$\Delta \xi$}&
\multicolumn{1}{c}{$\Delta$$T_{\rm eff}$}&
\multicolumn{1}{c}{$\Delta$log~$g$}&
\multicolumn{1}{c}{$\Delta \xi$} \\
& \multicolumn{1}{c}{$-200K$}&
\multicolumn{1}{c}{$+0.25$}&
\multicolumn{1}{c}{$+0.5$}&
\multicolumn{1}{c}{$-200K$}&
\multicolumn{1}{c}{$+0.25$}&
\multicolumn{1}{c}{$+0.5$}&
\multicolumn{1}{c}{$-200K$}&
\multicolumn{1}{c}{$+0.25$}&
\multicolumn{1}{c}{$+0.5$}&
\multicolumn{1}{c}{$-200K$}&
\multicolumn{1}{c}{$+0.25$}&
\multicolumn{1}{c}{$+0.5$} \\
\hline
C I & ...& ...& ...& $-$0.31& $-$0.07& $-$0.02 & $+$0.10& $+$0.01& $-$0.03 & $-$0.03& $+$0.06& $+$0.06 \\
N I & ...& ...& ...& ...& ...& ... & $+$0.19& $-$0.04& $-$0.02 & $-$0.27& $-$0.09& $+$0.04 \\
O I& $+$0.30& $-$0.03& $-$0.13& $+$0.03& $-$0.06& $-$0.03& $+$0.22& $-$0.05& $-$0.03& $-$0.24& $-$0.08& $+$0.01  \\
Na I& $+$0.21& $+$0.09& $-$0.01& $+$0.04& $+$0.07& $-$0.03& $-$0.03& $+$0.10& $-$0.03& $+$0.02& $+$0.13& $+$0.01 \\
Mg I& $+$0.18& $+$0.09& $+$0.06& $+$0.06& $+$0.06& $-$0.03& $-$0.08& $+$0.10& $+$0.06& $+$0.09& $+$0.12& $+$0.00 \\
Al I& $+$0.22& $+$0.07& $-$0.09 & $+$0.02& $+$0.07& $-$0.05& $-$0.04& $+$0.12& $-$0.05& $+$0.07& $+$0.13& $-$0.01 \\
Si I& $-$0.03& $+$0.02& $-$0.10& $+$0.01& $+$0.06& $-$0.04& $-$0.03& $+$0.10& $-$0.03& ...& ...&... \\
Si II& ...& ...& ...& ...& ...& ...& $+$0.17&$-$0.07&$+$0.09& $-$0.32& $-$0.11& $+$0.01 \\
S I& ...&...&...& $-$0.23& $-$0.02& $-$0.04&$+$0.03& $+$0.05& $-$0.03& $+$0.00& $+$0.10& $-$0.01 \\
Ca I& $+$0.24& $+$0.08& $+$0.03& $+$0.07& $+$0.06& $-$0.02& $-$0.07& $+$0.11& $+$0.01& ...&...&... \\
Ca II& ...& ...& ...& ...& ...& ...& $+$0.13& $-$0.04 & $-$0.04& ...& ...& ... \\
Sc II& ...& ...& ...& $-$0.44& $-$0.06& $-$0.01& $-$0.01& $-$0.09& $-$0.02& $-$0.01& $-$0.08& $-$0.01  \\
Ti I& $+$0.35& $+$0.06& $+$0.00& $+$0.28& $-$0.07& $+$0.01& ...& ...& ...&
...& ...& ... \\
Ti II& $+$0.14& $-$0.01& $+$0.01& $-$0.06& $-$0.19& $+$0.05& $+$0.00& $-$0.10& $+$0.07& $-$0.04& $-$0.10& $+$0.00 \\
Cr I& $+$0.25& $+$0.07&$+$0.36& $+$0.20& $+$0.08& $-$0.02& ...&...&...&...&
...& ... \\
Cr II& $-$0.08& $+$0.00& $-$0.04& $-$0.16& $+$0.03& $-$0.01& $+$0.06& $-$0.08& $+$0.00& $-$0.08& $-$0.10& $+$0.00 \\
Mn I& ...& ...& ...& $+$0.19& $+$0.03& $-$0.01& $-$0.07& $+$0.09&$-$0.02& ...&...&... \\
Co I& $+$0.00&$+$0.00&$+$0.10& ...&...&...& ...&...&...&...&...& ... \\
Ni I& $+$0.13& $+$0.02& $+$0.11& $+$0.13& $+$0.06& $-$0.04& $-$0.07& $+$0.10& $-$0.03& ...&...&... \\
Cu I& ...& ...& ...& $+$0.15& $+$0.06& $-$0.04& ...& ...& ...& ...& ...& ... \\
Zn I& ...& ...& ...& $+$0.06& $+$0.02& $+$0.02& $-$0.07& $+$0.10& $-$0.02& $+$0.06& $+$0.12& $-$0.01  \\
Y II& $+$0.18& $-$0.01& $+$0.00& $-$0.04& $-$0.04& $-$0.01& $-$0.02& $-$0.09& $+$0.01& ...&...&...  \\
Zr I& $+$0.43& $+$0.01& $+$0.03 & ...& ...& ...& ...& ...& ...& ...& ...& ... \\
Zr II& ...& ...& ...& ...& ...& ...&  $+$0.00& $-$0.08& $+$0.01 & ...& ...& ...\\
Mo I& $+$0.40& $+$0.04& $-$0.09 & ...& ...& ...& ...& ...& ...& ...&...& ... \\
Ba II& ...& ...& ...& $+$0.04& $-$0.08& $+$0.04 & ...& ...& ...&
...& ...& ... \\
La II& $+$0.29&$-$0.02&$-$0.08& ...& ...& ... & $-$0.05&$-$0.05&$-$0.02& ...& ...& ... \\
Ce II& $+$0.24&$-$0.02&$-$0.12& ...& ...& ...& $-$0.06&$-$0.06&$+$0.01& ...& ...& ...  \\
Pr II& $+$0.31&$-$0.02&$-$0.11& ...& ...& ...& $-$0.18&$-$0.02&$-$0.04& ...& ...& ... \\
Nd II& $+$0.26& $-$0.02&$-$0.08 & ...& ...& ...& $-$0.12& $-$0.03&$-$0.01 & ...& ...& ... \\
Sm II& $+$0.29&$-$0.01&$-$0.02& ...& ...& ...& $-$0.03&$-$0.05&$-$0.02& ...& ...& ...  \\
\hline 
\end{tabular}
\end{table*}
\subsection {\bf The errors caused by assumption of LTE}

 Non-LTE corrections for CNO elements
 show strong temperature dependence (particularly for N see e.g.
 Lyubimkov et al. 2011, Schiller \& Przybilla 2008) and they also
 vary from multiplet to multiplet. Venn (1995) have tabulated these
 corrections for C and N for a range of stellar temperatures for
 lines belonging to different multiplets. For B-F stars the correction
 for N varies from $-$0.3 dex to $-$1.0 dex while for C it is $-$0.1 to
 $-$0.5 dex. Takeda and Takada-Hidai (1998) have calculated
 Non-LTE corrections for oxygen abundance using 6156-6158\AA\ lines
 for A-F stars and correction varies between $-$0.1 to $-$0.4 dex.

 The neglect of departure from LTE also introduces errors
 in the estimated abundances of heavier elements.
 These errors for a given element vary
 with the stellar temperatures and also on metallicities. In very metal-poor
 stars the over-ionisation and in some cases usage of resonance lines
 for abundance determination results in errors as large as $+0.5$ dex for
 elements like Na. We have only used subordinates lines for sodium, believed
 to be formed in deeper layers and for them Non-LTE corrections of about
$-0.10$ dex are reported (Lind et al. 2011, Gehren et al. 2004).
 Gehren et al have estimated Non-LTE
corrections for Mg and Al and for the lines used in our analysis for a sample
of G stars covering a range in metallicities. For thick disc metallicities
Non-LTE corrections are $+0.05$ for [Mg/Fe] and $+$0.2 for [Al/Fe] is reported.
The Non-LTE corrections for S\,{\sc i} lines have been estimated by Korotin (2009).
 For the S\,{\sc i} lines used in our work (6042 $-$ 6056, 6743 $-$ 6758 \AA~), the effect is
 negligible. Wedemeyer (2001) has done Non-LTE calculation for Si\,{\sc i} lines
 and Non-LTE correction in range $-$0.01 to $-$0.05 have been estimated for
 the Sun and Vega respectively.
Mashonkina, Korn and Przybilla (2007) have studied departure from LTE
over a range of stellar parameter and meallicities for a large number of Ca\,{\sc i}  
and Ca\,{\sc ii} lines. At solar or moderately deficient metallicities the Non-LTE
correction is smaller than $+$0.1 dex for Ca\,{\sc i} lines used in
our analysis for deriving Ca abundance. Also the correction is negligible for
very weak Ca\,{\sc i} features.
Bergemann(2011) have computed Non-LTE Ti\,{\sc i} and Ti\,{\sc ii} for late type stars.
 At solar temperatures the Non-LTE corrections between $+$0.05 to 0.10 dex
 are found for Ti\,{\sc i} lines while the effect is negligible for Ti\,{\sc ii}  lines.
 Bergemann suggests the use of only Ti\,{\sc ii} lines and [Ti/Fe] computed from
 Ti\,{\sc ii} and Fe\,{\sc ii} would be more robust.
 Non-LTE analysis of Cr\,{\sc i} and Cr\,{\sc ii} by
Bergemann \& Cescutti (2010) for solar and metal-poor stars indicate
an error of $-$0.1 dex at the solar metallicities which becomes larger
$+$0.3 to 0.5 dex for very metal-poor stars.
 Non-LTE effects for Fe\,{\sc i} and Fe\,{\sc ii} lines have been estimated by
 Mashonkina (2011) for A-F stars with more complete representation of
 model atom. It is found that LTE underestimates the abundance derived
from Fe\,{\sc i} lines by 0.02 to 0.1 dex depending upon the chosen line
 and stellar temperatures; the effect could be as large as 0.2 dex
 for giants. The Non-LTE corrections are very small for Fe\,{\sc ii} lines.
 While deriving gravities from ionisation equilibrium of Fe\,{\sc i} \&
 Fe\,{\sc ii}, relatively large Non-LTE corrections for Fe\,{\sc i} must be taken
 into consideration.
 Velichko, Mashonkina \& Nilsson
(2010) have calculated Non-LTE correction for Zr\,{\sc i} and Zr\,{\sc ii} lines; both
 species become weak in LTE. The Non-LTE correction for Zr\,{\sc ii} lines
 used and for our stellar parameters range does not exceed $+$0.1 dex.
 Our sample stars are F-K supergiants with metallicity range 0 to $-$1.2.
 To disentangle the evolutionary effects (on abundances) from those
 inherent to the natal ISM we have compared the observed abundance pattern
 with those characterized for thin and thick disc by  extensive studies
 like  Bensby et al. 2005, Reddy et al. 2006 using extended samples of
 F-G dwarfs and subgiants and more recently by Takeda, Sato and Murata (2008)
 using late G giants. Between these samples the [X/Fe] vs [Fe/H] plots
 are remarkably similar for elements Si, Ca, Sc, Ti, V, Cr, Mn, Co, Ni and Cu
 and Non-LTE effects for these elements are also generally very small.
 Hence it is possible to infer the population type of the sample
 stars from important abundance ratios such as [$\alpha$/Fe] given for
 various disc components from the above mentioned studies.
 As explained above, the Non-LTE effects for $\alpha$ elements  
 are not large enough to mask the characteristic abundance differences.
\section{Results}

Results of our abundance analyses are presented in this section. If the
candidate PAGB star has preserved the composition of its AGB progenitor,
the C, N, O (and Li) abundances betray information about the
abundance changes brought by the first and second dredge-ups in the giant
prior to the AGB and changes brought about by the third dredge-up
on the AGB. {\bf We also look for the signature of Hot Bottom Burning (HBB) which can operate if the star had a massive progenitor (M$>$4M$_\odot$) (Ventura $\&$ D'Antona 2009) and its manifestation can produce enrichments of Li, N, Na, Mg and Al (Bl\"{o}cker $\&$ Sch\"{o}nberner 1991).}
 Abundances of elements {\bf from Si} through the iron group
are expected to be unchanged as a star evolves from the
main sequence to the AGB phase. These abundances may, in principle,
be used to assign a stellar population to the star, e.g., thin disc,
thick disc or halo. Beyond the iron group, abundances
of $s$-process nuclides are predicted to be enriched in AGB
stars that experienced the third dredge-up. Such nuclides dominate
elements such as Y, Zr and Ba.

\begin{table*}
\caption{References for the log {\itshape gf} values}
\label{table4}
\begin{tabular}{llllll}
\hline
\multicolumn{1}{l}{Species} & \multicolumn{1}{l}{Accuracy$^{a}$} & \multicolumn{1}{l}{Ref} &\multicolumn{1}{l}{Species} &\multicolumn{1}{l}{Accuracy$^{a}$} &\multicolumn{1}{l}{Ref} \\
\hline
C& B-C& 1& Fe\,{\sc ii} & A-B& 9 \\
N& B-C& 1& Co & B-C& 10 \\
O&  B-C& 2& Ni& B-D& 10 \\
Na&  A& 3 & Cu& B-D& 11 \\
Mg&  B-C& 4& Zn& B-D& 11 \\
Al&  B-D& 5& Y& B-D& 12\\
Si&  B-D&6& Zr& B-D& 13  \\
S&  D& 22 & Ba& B-D& 21 \\
Ca&  C-D& 23& La& B-D& 14 \\
Sc& D& 7& Ce& B-D& 15\\
Ti&  B-D &7& Pr& B-D& 16 \\
V&  B-D& 7& Nd& B-D& 17  \\
Cr&  B-C& 20& Sm& B-D & 18\\
Mn&  B-C& 7& Eu& B-D& 19\\
Fe\,{\sc i} &  A-B& 8& Dy& B-D& 16 \\
\hline
\end{tabular}
\flushleft$^{a}${Symbols indicating the accuracy of the {\itshape gf} values where
A = {$\pm$3$\%$}, B = {$\pm$10$\%$}, C = {$\pm$25$\%$} and D = {$\pm$50$\%$}}
\flushleft$^{1}${Wiese et al. 2007}, $^{2}${Wiese et al. 1996}, $^{3}${Sansonetti (2008)}, $^{4}${Kelleher et al. 2008a}
\flushleft$^{5}${Kelleher et al. 2008b}, $^{6}${Kelleher et al. 2008c}, $^{7}${Martin et al. 1988}, $^{8}${Fuhr et al. 2006}
\flushleft$^{9}${Mel\'{e}ndez \& Barbuy (2009)}, $^{10}${Fuhr et al. 1988}, $^{11}${Fuhr et al. 2005}, $^{12}${Hannaford et al. 1982} 
\flushleft$^{13}${Bi\'{e}mont et al. 1981}, $^{14}${Lawler et al. 2001a}, $^{15}$
{Lawler et al. 2009}
\flushleft$^{16}${Sneden et al. 2009}, $^{17}${Den Hartog et al. 2003}, $^{18}${Lawler et al. 2006}, $^{19}${Lawler et al. 2001b}, 
\flushleft$^{20}${Sobeck et al. 2007}, $^{21}${Curry (2004)}, $^{22}${Podobedova et al. 2009}, $^{23}${Aldenius et al. 2009}

\end{table*}

If the PAGB star has been subject to dust-gas winnowing or to
selective diffusion of elements according to the first ionization
potential (the FIP effect), abundance anomalies will be present
that are unrelated to those arising from nucleosynthesis.

In this section, we present an abundance analysis for each star
and comment briefly on the abundance anomalies, if any.
The primary initial goal is to draw attention to the
observational clues, if any, to the prior action of the
third dredge-up on the AGB and, thus, to the C, N, and O
abundances as well as the indicators of $s$-process contamination.
Another goal is to scan the anomalies for dust-gas winnowing and/or the
FIP effect.

We postpone to Section 5.13 a discussion of abundances, especially
the [X/Fe] values, which might betray the stellar population from
which the PAGB is drawn. One potential indicator of population is the
abundance of $\alpha$ elements Mg, Si, S, Ca and Ti which have a
higher ratio in thick disc stars than in the thin disc at the same [Fe/H].
This postponement arises in large part from the fact that in essentially
every star the [$\alpha$/Fe] ratios do not consistently point to a
thick or a thin disc origin. The $\alpha$-elements are sending a mixed
message whose interpretation is best discussed for the whole sample.

\begin{table*}
\begin{minipage}{140mm}
\centering
\caption{Elemental Abundances for IRAS 01259+6823 and IRAS 04535+3747.}
\label{table5}
\begin{tabular}{lllllllll}
\hline
&& \multicolumn{1}{l}{IRAS 01259+6823}&&&&
\multicolumn{1}{l}{IRAS 04535+3747}\\
\cline{3-5} \cline{7-9}  \\
\multicolumn{1}{l}{Species}&
\multicolumn{1}{l}{$\log \epsilon_{\odot}$}&
\multicolumn{1}{l}{[X/H]}& 
\multicolumn{1}{l}{N}&
\multicolumn{1}{l}{[X/Fe]}&
&\multicolumn{1}{l}{[X/H]}& 
\multicolumn{1}{l}{N}&
\multicolumn{1}{l}{[X/Fe]} \\
\hline
C I& 8.39 & $-0.42\pm 0.08$& 2& $+0.18$ && $-0.43\pm0.10$&5& $+0.05$
\\  
N I& 7.78 & ...&&&& $+0.24\pm0.08$&5& $+0.72$
\\
O I& 8.66 & $-0.29\pm 0.08$& 2& $+0.31$ && $-0.17\pm0.11$&3& $+0.31$
\\  
Na I& 6.17 & $-0.25\pm 0.05$& 3& $+0.35$&& $+0.05\pm0.04$&3&
$+0.53$ \\
Mg I& 7.53 & $-0.52\pm 0.10$& 2& $+0.08$&& $-0.46\pm0.14$& 4&
$+0.02$ \\
Al I& 6.37 & $-0.68\pm0.14$& 3& $-0.08$&& $-0.42\pm0.19$& 3& $+0.06$ \\
Si I& 7.51 & $-0.40\pm 0.09$& 10& $+0.20$&& $-0.27\pm0.14$&16&
$+0.21$ \\
Si II& 7.51 & $-0.19\pm 0.03$& 2& $+0.41$&&$-0.31\pm0.17$& 2&
$+0.17$ \\
S I& 7.14 & $-0.43$&synth& $+0.17$&&  $-0.16\pm0.15$& 2&
$+0.32$ \\
Ca I& 6.31 & $-0.79\pm0.09 $& 11& $-0.19$ && $-0.59\pm0.19$& 13&
$-0.11$  \\
Sc II& 3.05 & $-0.80\pm0.02$&synth&$-0.20$&&  $-0.75\pm0.00$& synth&
$-0.27$  \\
Ti I& 4.90 &$-0.67\pm 0.11 $& 11& $-0.07$&& $-0.56\pm 0.13$& 13& $-0.08$  \\
Ti II& 4.90 & $-0.70\pm 0.08 $& 3& $-0.10$ && $-0.70\pm0.08$& 5&
$-0.22$  \\
V I& 4.00 & $-0.80\pm 0.11 $& 11& $-0.20$&& ...  \\
Cr I& 5.64 & $-0.74\pm 0.07 $& 4& $-0.14$&&  $-0.58\pm0.12$& 8&
$-0.10$  \\
Cr II& 5.64 &$-0.52\pm 0.04 $& 4& $+0.08$&& $-0.57\pm0.09$& 5&
$-0.09$  \\
Mn  I& 5.39 &  $-0.72\pm0.08$&7&$-0.12$&& $-0.62\pm0.14$& 7&
$-0.14$  \\
Fe  & 7.45 & $-0.60$& & &&$-0.48$  \\
Co  I& 4.91 & $-0.75\pm0.11$& 3& $-0.15$&& ...  \\
Ni  I& 6.23 & $-0.79\pm 0.15 $& 21& $-0.19$ && $-0.53\pm0.14$&23&
$-0.05$   \\
Zn  I& 4.60 & $-0.69\pm 0.10 $& 3& $-0.09$ && $-0.53\pm0.13$& 3&
$-0.05$  \\
Sr I& 2.92 & $-0.30\pm 0.00 $& 1 & $+0.30$ && ...  \\
Y  II& 2.21 & $-0.53\pm0.11$& 5& $+0.07$ &&
$-0.40\pm0.20$&13 &$+0.08$  \\
Zr II& 2.58 & $-0.48\pm 0.02 $& 2 & $+0.12$ && $-0.49\pm0.17$&5 &$-0.01 $
 \\
Ba II& 2.17 & $-0.04\pm 0.00 $&synth&$+0.56$ && ...   \\
La II& 1.13 & $-0.19\pm 0.10 $& 3& $+0.41$ && $-0.33\pm0.23$& 2& $+0.15$  \\
Ce II& 1.58 & $-0.36\pm0.05$&8 &$+0.24$&& $-0.42\pm0.13$&18 &$+0.06$  \\
Pr II& 0.78 & $-0.48\pm0.06$& 3& $+0.12$ && $-0.31\pm0.23$&3 &$+0.17$  \\       
Nd II& 1.46 & $-0.38\pm0.07$& 10& $+0.22$ && $-0.42\pm0.08$&5 &$+0.06$  \\
Sm II& 0.95 & $-0.30\pm0.10$& 5& $+0.30$ && $-0.44\pm0.05$&3 &$+0.04$  \\
Eu II& 0.52 & $-0.31\pm0.01$& synth& $+0.29$ && ...  \\
Dy II&1.14 &...&&&& $-0.65\pm0.00$&2 &$-0.17$  \\
\hline
\end{tabular}
\end{minipage}
\end{table*}
   
\subsection{IRAS 01259+6823}

Our analysis appears to be the  first detailed abundance analysis
for this star with our final abundances presented in Table 5.
   A low resolution spectrum presented by Su\'{a}rez et al. (2006) shows
an absorption spectrum with emission limited to H$\alpha$.
   Stellar lines are sharp (FWHM about 0.26\AA\ corresponding to 14
 kms$^{-1}$) and symmetric at our resolution.

Atmospheric parameters (Table 2) estimated from Fe\,{\sc i} and Fe\,{\sc ii} lines
 were found to give satisfactory ionization equilibrium for Si, Ti, Cr and
 Fe: $\Delta$ = [X$_{II}$/H] $-$ [X$_{I}$/H]
 is $+$0.21, $-$0.03, $+$0.22 and $+$0.06 for  Si, Ti, Cr and Fe respectively.
The question of thick vs thin disc is discussed for the sample as whole in Section 5.13.

Of the elements anticipated to be affected by internal nucleosynthesis and
dredge-up by the termination of the AGB phase, C with [C/Fe] $= +0.2$ appears
enriched slightly by the third dredge-up in a AGB star; C is lowered by the
first dredge-up with [C/Fe] $\simeq -0.2$ .
The C/O ratio of $0.4$ by number shows that the star in not carbon-rich and,
hence, the third dredge-up was mild.
A mild $s$-process enrichment is seen
according to Y, Zr and Ba, with [$s$/Fe] $\simeq$ +0.3 with a slightly
higher enrichment for the so-called heavy $s$ elements (e.g., Ba) than for the
light $s$ elements (e.g., Y and Zr). Inspection of the abundances shows no sign for
operation of dust-gas winnowing or the FIP effect. In short, IRAS 01259+6823  evolved
at an early phase off the AGB before dredge-up from the He-shell began in earnest.

\subsection{IRAS 04535+3747}

 This is the first  abundance analysis
for this  SRD variable which with an effective temperature of 6000K has
presumably evolved from the AGB or possibly the horizontal branch. {\bf The star occupies the box VIa in the IRAS two colour diagram (Figure 1). This position is occupied by objects having detached cold CS dust shells.} 

Strong absorption lines have a weak
asymmetry in their blue wings. This asymmetry
has been ignored in the measurement of the equivalent
width.
We assume that a classical atmosphere may be used for the abundance
analysis. Model atmosphere parameters are given in Table 2 and the final abundances have been presented in Table 5.

The star is mildly metal-poor
 ([Fe/H] $= -0.48$).
Ionization equilibrium is adequately achieved not only for Fe
but also
for Si, Ti and Cr: $\Delta$ = [X$_{II}$/H] $-$ [X$_{I}$/H]
is $-0.04, -0.14, +0.01$ and $+0.03$  for Si, Ti, Cr and Fe respectively.

This star has not evolved from an AGB star in which the third dredge-up
has enriched the atmosphere in C and the $s$-process. The C/O ratio of 0.3 by number
is that expected of a star at the beginning of the AGB and the $s$-process has not
enriched the heavy elements. To within the uncertainities
about the initial C, N and O abundances, their alterations by the first
dredge-up and the errors of the abundance analysis, the C, N and O
abundances are those anticipated for a star beginning the AGB {\bf following either a stay
on the horizontal branch or may execute blue loops given the present uncertainities in the masses of SRD's.} 

\subsection{IRAS 05208-2035}

The infrared excess is
typical of RV Tauri variables (De Ruyter et al. 2006).
De Ruyter et al. complain lack of a published spectrum and
of estimates for the stellar parameters. This lack we correct here.
 An orbital period of 236 days has been reported in Gielen et al. (2008).

The chosen model (Table 2)
gives the abundances as presented in Table 6.
Ionization equilibrium is satisfactorily accounted for
in that $\Delta$ = [X$_{II}$/H] $-$ [X$_{I}$/H]
is $+0.01$, $-0.02$, and $+0.05$ for Fe, Ti and Cr respectively.
In terms of the evolutionary history, the most significant aspect of
the star's composition is an enhancement of the heavy elements. In particular, those
elements (e.g., Y, Zr and Ba) dominated by an $s$-process contribution are
overabundant relative to Fe by about 0.3 dex. The $r$-process dominated
elements (e.g., Eu) are also overabundant but this is as expected for a star of
[Fe/H] $= -0.6$.
 Although the star is a spectroscopic
binary and thus a candidate for dust-gas winnowing, this effect is absent but this is
not surprising given the extensive convective envelope of such a cool star.

\subsection{IRAS 07140-2321}

IRAS 07140-2321 (also known as SAO 173329) is a
 PAGB star belonging to a spectroscopic binary with a dusty circumbinary
disc (Van Winckel et al. 2000, De Ruyter et al. 2006; Gielen et al. 2008).
The orbital period is 116 days (De Ruyter et al. 2006).
Photometrically, the star is an irregular small amplitude
variable for which Kiss et al. (2007) suggests variations with periods of
about 24 and 60 days.
Here, we are responding to a call by Van Winckel (1997) who published the only previously
reported abundance analysis and wrote `SAO 173329 is a metal-deficient
object for which more data are needed'. Specifically, Van Winckel noted the lack of
an oxygen abundance determination and the need for `more good Fe\,{\sc ii} lines'.
The lack of abundance data for heavy or $s$-process elements was also a notable
omission.

Results from our analysis summarized in Table 6 are based on the model atmosphere
summarized in Table 2. The star is metal-poor: [Fe/H] $= -0.9$.
The C/O ratio of 0.4 and the limited data on heavy element abundances show that this
star did not evolve from a thermally pulsing AGB star. The N abundance shows evidence
of N enrichment by the First Dredge-Up (FDU). The observed [N/Fe] after Non-LTE
correction of $-0.4$ dex (Lyubimkov et al. 2011) is $+0.3$ dex which is similar to
 FDU prediction of $+$0.5 dex
as given in Schaller (1992). 
Heavy elements -- Y, Zr and Ba are not
enriched relative to Fe.
For elements in common, our abundances are in fair agreement with those
reported by  Van Winckel (1997) who used
solar {\itshape gf}-values and  a different grid
 of model atmospheres.
Abundance differences between those in Table 6 and those reported by
Van Winckel for a 7000 K model atmosphere are small (except for S\,{\sc i} and Cr\,{\sc i}):
  $\delta$ [X/Fe] (present work $-$ VW97) is $-$0.20 dex
 for C\,{\sc i}, $+$0.11 dex for N\,{\sc i}, $+$0.43 dex for S\,{\sc i}, $-$0.22 for
 Ca\,{\sc i}, $-$0.22 for Ti\,{\sc ii}, $-$0.37 for Cr\,{\sc i}, $+$0.14 for Mn\,{\sc i},
 $+$0.13 for Ni\,{\sc i} and $-$0.03 for Zn\,{\sc i}.

The inspection of the estimated abundances shows that the observed [S/Fe] is $+$0.6;
the possible $\alpha$ enrichment of $+$0.3 dex expected at [Fe/H] of $-$0.9
may indicate actual [S/Fe] of $+$0.3 dex. But nearly zero [Zn/Fe] and lack of depletion
 for high T$_{C}$ elements Ca and Sc
 shows that the star is not affected by dust-gas winnowing. It is surprising
given the fact that the star is a spectroscopic binary with an orbital period of
only 115.9 days (Van Winckel $\&$ Reyniers 2000) and has detected circumstellar material.
It appears that conditions for effective dust-gas winnowing are far from understood.

\subsection{IRAS 07331+0021}

 This cool variable exhibits TiO bands in its spectrum at its coolest
phases
(Klochkova \& Panchuk 1996) which
necessarily impair a full abundance analysis at such phases. Luck \& Bond (1989) 
undertook an analysis of an image-tube spectrum obtained in 1981
and found the star was metal-poor ([Fe/H]$=-1.0$) with a
larger under-abundance of $s$-process elements. Noting that all of the
elements deficient with respect to iron had second ionization potentials
less than the ionization potential of hydrogen, they speculated that
Lyman continuum emission from shock waves in the atmosphere over-ionized
these elements and therefore they appear under-abundant when a standard
analysis with a classical atmosphere is performed. Klochkova \& Panchuk (1996, 1998)
reported abundance analyses from CCD spectra of AI CMi taken at three different
epochs. Results for the separate phases are given in their 1996 paper and
average results with the addition of O and Zn are provided in the 1998
paper. Inspection of the 1996 tabulation of abundances shows that results for
several species vary from spectrum to spectrum: for example, the [Sc/Fe] ratio
is variously given as $-0.63$, $-0.22$, and $-0.05$. Luck \& Bond's shocking
hypothesis might allow for this kind of variation. From their comparison with
Luck \& Bond's results, Klochkova \& Panchuk (1996) conclude that the two
analyses are in good agreement. They further write 'We come to the conclusion that from
1981 to 1995 abundances of elements heavier than oxygen remained the same in the
AI CMi atmosphere, i.e., no transfers of matter processed in the inner layers of the
star to the star's atmosphere is observed.'

 Our spectrum taken on April 19, 2008 exhibits strong bands of CN, CH, MgH and TiO.
 Strong lines exhibit blue-shifted components.
 The atmospheric parameters (Table 2) fall within the range
reported by Klochkova and Panchuk (1998).
 We estimate [Fe/H] of $-$1.16 while KP1998 derive $-$1.14.
 For the most elements the abundances derived by us agree with KP1998.
\begin{table*}
\centering
\caption{Elemental Abundances for IRAS 05208-2035, IRAS 07140-2321 and IRAS 07331+0021}
\label{table6}
\begin{tabular}{lllllllllllll}
\hline
&& \multicolumn{1}{l}{IRAS 05208-2035}&&&&
\multicolumn{1}{l}{IRAS 07140-2321} &&&& \multicolumn{1}{l}{IRAS 07331+0021}\\
\cline{3-5} \cline{7-9} \cline{11-13}  \\
\multicolumn{1}{l}{Species}&
\multicolumn{1}{l}{$\log \epsilon_{\odot}$}&
\multicolumn{1}{l}{[X/H]}&
\multicolumn{1}{l}{N}&
\multicolumn{1}{l}{[X/Fe]}&
&\multicolumn{1}{l}{[X/H]}&
\multicolumn{1}{l}{N}&
\multicolumn{1}{l}{[X/Fe]}
&& \multicolumn{1}{l}{[X/H]}&
\multicolumn{1}{l}{N}&
\multicolumn{1}{l}{[X/Fe]}
 \\
\hline

C I& 8.39 &....&&&& $-0.61\pm0.10$& 12&
$+0.31$&& $-0.44\pm0.03$& 2&$+0.72$  \\
N I& 7.78 &....&&&& $-0.15\pm0.16$& 4& $+0.77$ && ...  \\
O I& 8.66 &$-0.25\pm0.27$& 2& $+0.40$ && $-0.54\pm0.00$& 1& $+0.38$&
& $+0.04\pm0.09$& 2& $+1.20$\\
Na I& 6.17 & $-0.22\pm0.09$& 5 & $+0.43$ & & $-0.40\pm0.16$& 4&
$+0.52$ & & ...\\
Mg I& 7.53 & $-0.52\pm0.14$& 2& $+0.13$ & & $-0.61\pm0.14$& 3&
$+0.31$ & & $-0.68\pm0.00$& 1&$+0.48$ \\
Mg II& 7.53& ...&&&& $-0.74\pm0.00$& 1&$+0.18$ && ... \\
Al I& 6.37 & $-0.45\pm0.05$& 4& $+0.20$ & & $-1.18\pm0.00$& 1&$-0.26$ && .... \\
Si I& 7.51 & $-0.56\pm0.19$&8& $+0.09$ & & $-0.56\pm0.12$&5&
$+0.36$&  & $-0.89\pm0.24$&5& $+0.27$ \\
Si II& 7.51 &...&&&& $-0.74\pm0.00$& 1& $+0.18$ && ... \\
S I&   7.14 &...&&&& $-0.32\pm0.04$& 3 &$ +0.60$ && ... \\
Ca I& 6.31 & $-0.79\pm0.13$& 6 & $-0.14$ && $-0.97\pm0.19$& 15&
$-0.05$ & & $-1.25\pm0.09$& 3& $-0.09$\\
Sc II& 3.05 & ... &&&& $-0.92\pm0.13$& 9&
$+0.00$ & & $-1.35\pm0.15$& 2& $-0.19$\\
Ti I& 4.90 & $-0.52\pm0.24$& 15& $+0.13$ & &....&&&&  ... \\
Ti II& 4.90 & $-0.54\pm0.26$& 4& $+0.11$&&  $-0.73\pm0.14$& 8&$+0.19$ &&
$-0.89\pm0.05$&5& $+0.27$  \\
Cr I& 5.64 & $-0.64\pm0.12$& 6&$+0.01$& & $-0.71\pm0.10$& 4& $+0.21$&
& $-0.98\pm0.16$& 2& $+0.18$\\
Cr II& 5.64 &$-0.59\pm0.00$& 2& $+0.06$& & $-0.74\pm0.05$& 5& $+0.18$
& & $-0.97\pm0.10$& 2& $+0.19$ \\
Mn  I& 5.39 & ...& &&& $-1.00\pm0.01$& synth &$-0.08$ &&
$-1.35\pm0.08$& 3& $-0.19$ \\
Fe   & 7.45 & $-0.65$& & &&$-0.92$ &&&&
$-1.16$ & & \\
Co  I& 4.92 & $-0.53\pm 0.08$&2& $+0.12$ &&....&&&& $-0.97\pm0.11$& 6& $+0.19$ \\
Ni  I& 6.23 & $-0.44\pm0.17$& 13& $+0.21$& & $-0.85\pm0.07$& 7&
$+0.07$ & & $-1.19\pm0.10$&10& $-0.03$ \\
Cu I& 4.21 & ...&&&& ...&&&& $-1.30\pm0.08$& synth& $-0.14$ \\
Zn  I& 4.60 &...&&&& $-0.87\pm0.16$& 2 &$+0.05$ && ... \\
Y  II& 2.21 & $-0.41\pm0.16$& 4& $+0.24$ && $-1.10\pm0.13$& 5&
$-0.18$ & & $-1.15\pm0.00$& 1& $+0.01$\\
Zr I& 2.59 & $-0.30\pm0.10$& 4& $+0.35$ &&....&&&&... \\
Mo I& 1.92 & $-0.16\pm0.12$& 2& $+0.49$ &&...&&&&... \\
Ba II& 2.17 & $-0.34\pm0.00$& synth& $+0.31$ && $-0.64\pm0.00$& synth &$+0.28 $ &&
... \\
La II& 1.13 & $-0.31\pm0.15$& 3& $+0.34$ &&...&&&& $-0.89\pm0.11$& 3& $+0.27$ \\
Ce II& 1.58 & $-0.29\pm0.05$& 2& $+0.36$ &&...&&&& $-1.22\pm0.00$& 1& $-0.06$ \\
Pr II& 0.71 & $-0.40\pm0.00$& 1 & $+0.25$ &&...&&&& $-1.02\pm0.00$& 1& $+0.14$ \\
Nd II& 1.45 & $-0.31\pm0.11$& 6 & $+0.34$ &&...&&&& $-1.36\pm0.22$& 2& $-0.20$ \\
Sm II& 1.01 & $-0.39\pm0.18$& 3 & $+0.26$ &&...&&&& $-1.02\pm0.13$& 5& $+0.14$ \\
Eu II& 0.52 & $-0.51\pm0.03$& synth & $+0.14$ &&...&&&& .... \\
\hline
\end{tabular}
\end{table*}
 Our analysis covers additional elements C, Co and Sm (see Table 6).
Of note is that our analysis and that of KP1998 do not confirm the
under-abundance of heavy elements reported by Luck \& Bond. Heavy elements
have a normal abundance for a star with [Fe/H] $= -1.16$. The C and especially
the O
abundance are unusually high.  C abundance has been measured
 from 5380.3 and O from [O\,{\sc i}] lines. 
 A fair conclusion may be that the star like others in
our sample has not experienced the full effect of the third dredge-up on the
AGB branch.

\subsection{IRAS 08187-1905}

High-resolution optical spectra have not been previously reported.
Low
resolution spectroscopy and photometry was obtained by (Reddy $\&$ Parthasarathy 1996).
Our abundance analysis (Table 7) with the model atmosphere in
Table 2 suggests that the star is slightly metal-poor ([Fe/H] $= -0.6$) and is
C-rich. The O abundance is as expected for a [Fe/H]$= -0.6$ star but the C
enrichment suggests addition of C to the original material. Formally, C/O
by number of atoms slightly exceeds unity. The light-$s$ elements Y and Zr as well as the heavy $s$ element Ba point to
a mild $s$-process enrichment.
One is led to suggest that the star, after a FDU which decreased the
C and increased the N abundance, experienced the third dredge-up which increased the
C abundance and thus the C/O ratio and also mildly enriched the $s$-process
content of the atmosphere.

\subsection{HD 107369}

This high-galactic latitude F supergiant is not an IRAS source. Van Winckel (1997)
who undertook an abundance analysis using high-resolution spectra of
selected wavelength intervals found the star to be metal-poor ([Fe/H] $= -1.3$).
Our abundance analysis with the model in Table 2 is reported in Table 7.
 We estimate a temperature of 7500K
 and log~$g$ of 1.5 using the Fe\,{\sc i} and Fe\,{\sc ii} lines. However,
 the estimated gravity is also supported by ionization equilibrium of
  Mg\,{\sc i} \& Mg\,{\sc ii}, Ca\,{\sc i} \& Ca\,{\sc ii}, Cr\,{\sc i} \& Cr\,{\sc ii}.
Our abundances are in good agreement with Van Winckel's results.
To make comparison easy we have used his line strengths with new set of {\itshape gf} values
  and models. The scanty Fe line data supports the temperature and gravity
  estimated by us. Adopting our model (7500,1.5,2.2) we find following
  $\delta$ (present work $-$ VW1997)
  $+$0.34 dex for N\,{\sc i}, $+$0.13 dex for O\,{\sc i}, $+$0.18 for
 Si\,{\sc ii}, $+$0.24 for S\,{\sc i}, $-$0.09 for Ca\,{\sc i}, $+$0.26 for Ti\,{\sc ii}, $-$0.03 for Cr\,{\sc ii},
 $+$0.06 for Fe\,{\sc i}, $+$0.12 for Fe\,{\sc ii} and $+$0.04 for Ba\,{\sc ii} respectively.
 Our analysis covers additional elements Mg, Ni, Sr and Y.
The star does not provide the expected abundance signature of a typical AGB star
  such as C/O greater than one and a strong $s$-process enrichment. Nor do we see
 any indication of dust-gas winnowing. The lines of Zn were not detectable.

\begin{table*}
\centering
\caption{Elemental Abundances for IRAS 08187-1905, HD 107369 and IRAS 12538-2611}
\label{table7}
\begin{tabular}{lllllllllllll}
\hline
&& \multicolumn{1}{l}{IRAS 08187-1905}&&&&
\multicolumn{1}{l}{HD 107369} &&&& \multicolumn{1}{l}{IRAS 12538-2611}\\
\cline{3-5} \cline{7-9} \cline{11-13}  \\
\multicolumn{1}{l}{Species}&
\multicolumn{1}{l}{$\log \epsilon_{\odot}$}&
\multicolumn{1}{l}{[X/H]}&
\multicolumn{1}{l}{N}&
\multicolumn{1}{l}{[X/Fe]}&
&\multicolumn{1}{l}{[X/H]}&
\multicolumn{1}{l}{N}&
\multicolumn{1}{l}{[X/Fe]}
&& \multicolumn{1}{l}{[X/H]}&
\multicolumn{1}{l}{N}&
\multicolumn{1}{l}{[X/Fe]}
 \\
\hline
C I& 8.39 & $+0.03\pm0.25$& 9& $+0.62$ && ...
&&&& $-1.09\pm0.13$& 3& $+0.03$ \\
N I& 7.78 &$-0.10\pm0.23$& 4&$+0.49$&&  $-0.18\pm0.25$& 5& $+1.15$ &&
... \\
O I& 8.66 &$-0.33\pm0.24$& 3& $+0.26$ &&  $-0.68\pm0.09$& 4& $+0.65$&
& $-0.17\pm0.01$& 2& $+0.95$ \\
Na I& 6.17 & $+0.03\pm0.19$& 3 & $+0.62$ && ... &&&& $-0.61\pm0.05$& 3& $+0.51$ \\
Mg I& 7.53 & $-0.46\pm0.19$& 3& $+0.13$ & & $-0.95\pm0.04$& 2&
$+0.38$ && $-0.80\pm0.00$& 1& $+0.32$ \\
Mg II& 7.53& ...&&&& $-0.77\pm0.04$& 2&$+0.56$ && ... \\
Al I& 6.37 &...&&&& ...&&&& $-1.38\pm0.11$& 2& $-0.26$ \\
Si I& 7.51 & $-0.11\pm0.15$&5& $+0.48$ & & ...
&&&& $-0.89\pm0.19$& 6& $+0.23$ \\
Si II& 7.51 & ...&&&& $-0.80\pm0.24$& 4& $+0.53$
 && ... \\
S I& 7.14 & $+0.05\pm0.06$& 3 &$+0.64$&& $-0.84$&synth&$+0.49$&& $-0.69
\pm0.01$& 2& $+0.43$ \\
Ca I& 6.31 & $-0.63\pm0.17$& 8 & $-0.05$ && $-1.28\pm0.09$& 3&
$+0.05$ & & $-1.34\pm0.14$& 10& $-0.22$ \\
Ca II& 6.31&....&&&& $-1.43\pm0.00$&1 & $-0.10$ && ... \\
Sc II& 3.05 & $-0.75\pm0.00$& synth & $-0.16$ && $-1.45$& synth&
$-0.12$ & & $-1.55\pm0.03$& synth& $-0.43$ \\
Ti I& 4.90 &....&&&&.....&&&& $-0.81\pm0.00$& 1&$+0.31$  \\
Ti II& 4.90 & $-0.99\pm0.10$& 2& $-0.40$&&  $-1.09\pm0.10$& 18&$+0.24$ && $-0.89\pm0.19$& 6& $+0.23$ \\
Cr I& 5.64 & $-0.52\pm0.15$& 3&$+0.07$& & $-1.52\pm0.05$& 2& $-0.19$&& $
-1.50\pm0.02$& 2&$-0.38$ \\
Cr II& 5.64 & $-0.66\pm0.14$& 7&$-0.08$& & $-1.43\pm0.10$& 7& $-0.10$&& $-1.49\pm0.05$& 4& $-0.37$ \\
Mn  I& 5.39 & ...  &&&& ...&&&& $-1.52\pm0.06$& 2&$-0.40$ \\
Fe & 7.45 & $-0.59 $& & &&$-1.33 $ &&&&$-1.12$ & & \\
Ni  I& 6.23 & $-0.60\pm0.11$& 7& $-0.01$& & $-1.05\pm0.04$& 2&
$+0.28$ & & $-1.13\pm0.18$& 11& $-0.01$ \\
Zn  I& 4.60 & $-0.58\pm0.16$& 3& $+0.01$ && ... &&&& $
-1.04\pm0.06$& 2&$+0.08$ \\
Sr II& 2.92& ...&&&& $-1.85\pm0.00$&1&$-0.52$ && ... \\
Y  II& 2.21 & $-0.19\pm0.20$& 4& $+0.40$ && $-1.81\pm0.07$& 2&$-0.48$ && $-1.28\pm0.09$& 2& $-0.16$ \\
Zr II & 2.59&$-0.33\pm0.14$&2&$+0.25$ &&...&&&& ... \\
Ba II& 2.17 & $-0.34\pm0.00$& synth& $+0.25$ && $-1.80\pm0.12$& 2& $-0.47$ && $
-1.34\pm0.00$& 1&$-0.22$ \\
La II& 1.13& ...&&&& ...&&&& $-1.37\pm0.05$& 2& $-0.25$ \\
Ce II& 1.58& ...&&&& ...&&&& $-1.45\pm0.07$& 3& $-0.33$ \\
Nd II& 1.45& ...&&&& ...&&&& $-1.51\pm0.07$& 3& $-0.39$ \\ 
\hline
\end{tabular}
\end{table*}

 The elements with higher condensation temperature such as Ca, Sc did not exhibit expected depletion.
 The N abundance appears very high at [N/Fe] = $+$1.15 and in excess of
the value expected by the FDU. However, Non-LTE corrections for the
N\,{\sc i} lines are $-$0.5 dex from Lyubimkov et al. 2011 is indicated. After applying this correction
we get [N/Fe] of $+$0.7 dex which is marginally larger than prediction of FDU.   
Van Winckel reported [C/Fe] of $-$0.2 dex from a single line.
 There is no compelling evidence for HBB in the form
of very large [N/Fe] nor do we see Li\,{\sc i} feature. {\bf Being a high galactic latitude
 metal-poor object it does not seem to have evolved from a massive progenitor since
 M $>$ 4M$_{\odot}$ is required for HBB (Ventura \& D'Antona 2009).} With low C/O ratio, high galactic latitude
 and low metallicity it appears to have evolved from a low mass progenitor.

\subsection{IRAS 12538-2611}

 Also known as HR 4912, this variable star
 was studied spectroscopically by Luck, Lambert and Bond (1983)
 (LLB83) who found it metal-poor ([Fe/H]=$-$1.2). With its high
 luminosity it is quite likely to be a PAGB object.
   An abundance analysis was also performed by
  Giridhar, Arellano Ferro and Parrao (1997) (GAFP97).
 Our spectra have better resolution and spectral coverage than earlier studies.
There is fair agreement between our abundances and those by LLB83 and GAFP97.
The run of C, N and O abundances offer no evidence that the star evolved
from a thermally pulsing AGB star: the C/O ratio is 0.06 by number of atoms (see Table 7). The N
abundance (estimated by LLB83) before correction for Non-LTE effects is consistent
with that expected from the FDU. There is no evidence for
$s$-process enrichment.

\subsection{IRAS 17279-1119}

This PAGB star has been
the subject of two previous abundance analyses. In introducing IRAS 17279-1119, Van Winckel (1997) hereinafter VW97
noted that optical photometry shows that it has some RV Tauri-like characteristics with
possible periods of 61 and 93 days (Bogaert 1994) but additional photometry is needed;
inspection of SIMBAD shows that photometry has not been reported recently.
Van Winckel's abundance analysis for a model atmosphere with {\itshape $T_{\rm eff}$}= 7400K and
{\itshape log g}=0.5 showed the star to be metal-poor ([Fe/H] $=-0.7$)
with a C/O close to unity and possibly exceeding unity and
$s$-process enriched, both markers of prior third dredge-up on the AGB. His discussion
of the star concluded with the observation that ``this as yet poorly studied star deserves
further research'. An appreciated limitation of Van Winckel's analysis was the
limited wavelength coverage of his spectroscopic snapshots.
Arellano Ferro, Giridhar $\&$ Mathias (2001) hereinafter AGM2001
acquired new spectra with complete coverage from 3900-6800 \AA\ which enabled them to determine
the abundances of all elements considered by VW97 (except for N) and add Na, Mg, Mn, La
and Ce to the list. The star was again shown to be metal-poor ([Fe/H]$-0.6$)
and $s$-process enriched. Unusual aspects of the composition are found from these analyses.
Most striking perhaps, AGM2001 found [Na/Fe] =$+$0.6 and from both analyses a high
Sc abundance was reported, [Sc/Fe]=$+$0.9 (VW97 from one Sc\,{\sc ii} line) and
$+$0.6 (AGM2001 from eight Sc\,{\sc ii} lines).

 Since the solar abundances used by AGM2001 and VW97 are different from our work we have transformed these abundances to the abundance scale
 of Asplund et al. (2005) for comparison.
 The present analysis covers additional elements Sr, Zr, Pr and Nd (see Table 8).
In general, the results of the present analysis agree well with those of prior
analyses. An exception is
\begin{table*}
\centering
\caption{Elemental Abundances for IRAS 17279-1119, IRAS 22223+4327 and BD +39 4926}
\label{table8}
\begin{tabular}{lllllllllllll}
\hline
&& \multicolumn{1}{l}{IRAS 17279-1119}&&&&
\multicolumn{1}{l}{IRAS 22223+4327} &&&& \multicolumn{1}{l}{BD +39 4926}\\
\cline{3-5} \cline{7-9} \cline{11-13}  \\
\multicolumn{1}{l}{Species}&
\multicolumn{1}{l}{$\log \epsilon_{\odot}$}&
\multicolumn{1}{l}{[X/H]}&
\multicolumn{1}{l}{N}&
\multicolumn{1}{l}{[X/Fe]}&
&\multicolumn{1}{l}{[X/H]}&
\multicolumn{1}{l}{N}&
\multicolumn{1}{l}{[X/Fe]}
&& \multicolumn{1}{l}{[X/H]}&
\multicolumn{1}{l}{N}&
\multicolumn{1}{l}{[X/Fe]}
 \\
\hline

C I& 8.39 & $-0.08\pm0.17$& 11 &$+0.35$ && $+0.27\pm0.14$& 20&$+0.60$ && $+0.05\pm0.12$& 17 & $+2.42$ \\
N I& 7.78 & $+0.20\pm0.12$& 4&$+0.63$ && $+0.33\pm0.01$& 2&$+0.66$ && $+0.52\pm0.26$& 9& $+2.89$ \\
O I& 8.66 &$-0.31\pm0.00$& 1& $+0.12$ &&  $+0.07\pm0.05$& 2&$+0.40$ && $+0.19\pm0.09$& 10& $+2.56$ \\
Na I& 6.17 &$-0.04\pm0.12$& 2&$+0.39$ && $+0.10\pm0.06$& 2&$+0.43$ && $-0.39\pm0.17$& 2& $+1.98$ \\
Mg I& 7.53 & $-0.42\pm0.10$& 3& $+0.01$ & & $+0.07\pm0.09$& 2&
$+0.40$ && $-1.91\pm0.15$& 4& $+0.46$ \\
Al I& 6.37 & ... &&&& $+0.15\pm0.01$& 2&$+0.48$ && $-2.79\pm0.00$& 1& $-0.42$ \\ 
Si I& 7.51 & $-0.19\pm0.22$&4& $+0.24$ & & $+0.09\pm0.07$& 4&$+0.42$&&..&& \\
Si II& 7.51 & $-0.24\pm0.09$& 2&$+0.19$ && $+0.06\pm0.00$&1&$+0.39$ && $-1.87\pm0.25$& 6& $+0.50$ \\
S I& 7.14 & $+0.03\pm0.12$& 4 & $+0.46$ && $+0.25\pm0.10$& 2&$+0.58$ && $+0.07\pm0.17$& 4& $+2.44$ \\
Ca I& 6.31 & $-0.54\pm0.19$& 7 & $-0.11$ & & $-0.20\pm0.09$& 8&$+0.13$ &&.. \\
Ca II& 6.31&...&&&& $-0.22\pm0.08$& 2 &$+0.11$ &&... \\
Sc II& 3.05 & $-0.43\pm0.03$& synth & $+0.00$ && $-0.38\pm0.05$& 2&
$-0.05$ && $-2.80\pm0.08$& 2& $-0.43$ \\
Ti II& 4.90 & $+0.04\pm0.22$& 6& $+0.47$&&  $-0.17\pm0.08$& 6&$+0.16$ && $-2.73\pm0.14$& 10& $-0.36$ \\
Cr II& 5.64 & $-0.53\pm0.02$& 3&$-0.09$ && $-0.41\pm0.15$&8&$-0.08$ && $-2.07\pm0.15$& 5& $+0.30$ \\
Mn  I& 5.39 & $-0.62\pm0.27$& 2& $-0.19$ && $-0.46\pm0.05$& 2&$-0.13$ &&... \\
Fe   & 7.45 & $-0.43$& & &&$-0.33 $ &&&& $-2.37$ \\
Ni I& 6.23 & $-0.60\pm0.43$& 2& $-0.17$ &&  $-0.22\pm0.11$& 11&$+0.11$ &&... \\
Zn I& 4.60 & $-0.28\pm0.11$& 2& $+0.15$ && $-0.45\pm0.00$& 1&$-0.12$ && $-0.70\pm0.08$& 2& $+1.67$ \\
Sr I& 2.92& $-0.20\pm0.00$& 1& $+0.23$ &&...&&&&... \\
Y  II& 2.21 & $+0.50\pm0.22$& 5& $+0.93$ && $+1.15\pm0.15$& 3&$+1.48$ &&... \\ 
Zr II& 2.58 & $+0.52\pm0.07$& 2& $+0.95$ && $+0.82\pm0.16$& 5&$ +1.15$ &&.. \\
Ba II& 2.17 & $+0.16\pm0.00$& synth& $+0.59$ && $+1.16\pm0.00$& synth&$+1.49$ &&.. \\
La II& 1.13 &$+0.18\pm0.17$& 2& $+0.61$ && $+0.97\pm0.14$& 16& $+1.30$&&... \\ 
Ce II& 1.58 & $+0.43\pm0.11$&7&$+0.86$ && $+0.48\pm0.13$& 11&$+0.81$&&.. \\
Pr II& 0.78 & $+0.00\pm0.00$& 1& $+0.43$ && $+0.70\pm0.09$& 5&$ +1.03$&&..\\
Nd II& 1.45 & $+0.26\pm0.06$&2&$+0.69$ && $+0.60\pm0.10$& 14&$ +0.93$&&.. \\
Sm II& 0.95 & ...&&&& $+0.50\pm0.07$& 5&$+0.83$ && ... \\
Eu II & 0.52 &  ...&&&& $+0.10\pm0.01$&synth&$+0.43$ &&... \\
\hline
\end{tabular}
\end{table*}
 that we do not confirm the anomalously high Sc abundance
reported by VW97 and AGM2001.

 The progenitor of IRAS 17279-1119 was likely a thermally pulsing AGB star. This suggestion
accounts for the C/O of unity and the modest $s$-process enrichment.

\subsection{IRAS 22223+4327}

 This star has been classified as a proto-planetary nebula  and, therefore, a
PAGB star (Hrivnak 1995, Kwok 1993).
  Even on a low resolution spectrum, Hrivnak (1995) could see the
  enhancement of lines of s-process elements.
   From the UBV photometry, Arkhipova et al. (2003)
  found this star to be pulsating variable with a period of $\sim$ 90 days.

  Long term  monitoring of this star using high resolution
  spectra has been carried out by Klochkova, Panchuck \& Tavolzhanskaya (2010)
  with following interesting findings.
   The strong absorption lines such as low excitation line of Ba\,{\sc ii} at
   6141 \AA\  not only show asymmetries in the profile with short wavelength
    side of the profile showing extended wing than the red wing;  these strong lines also
  show large amplitude profile variations (with time) caused by
  variations in blue wing while red wing remained unchanged.
  The spectrum contains C$_{2}$ lines most likely formed in the
 circumstellar shell. At the epoch of largest asymmetry in strong Ba\,{\sc ii} 
  line the C$_{2}$ (0;1) band head at $\lambda$ 5635 is seen in emission.
  The cores of hydrogen lines show larger variations in radial velocities
  ($\sim$ 8 kms$^{-1}$) while weak metallic lines show smaller amplitude
 variations in radial velocities ($\sim$ 1 kms$^{-1}$). Molecular C$_{2}$ lines
 remain stationary with time; the shift in circumstellar features relative
 to systemic velocity gives an expansion velocity V$_{exp}$ of 15.0 kms$^{-1}$.
   Our spectrum taken on Dec 27, 2009 also exhibits the features mentioned in
   Klochkova, Panchuck \& Tavolzhanskaya (2010).
 This star was analysed by
 Van Winckel and Reyniers (2000) (hereinafter WR2000) who found it moderately metal-poor
 [Fe/H] of $-$0.3 dex and showing enhancement of s-process elements.
  The present analysis covers additional elements Na, Mg and Zn and uses larger number of
 lines for many species (see Table 8).
  Since the solar abundances used in WR2000 are different from our work, we have
  transformed these abundances to solar abundances of Asplund et al. (2005) to facilitate
  comparison. All elements agree within $\pm$0.15 dex.

\begin{figure}
\begin{center}
\includegraphics[width=9cm,height=9cm]{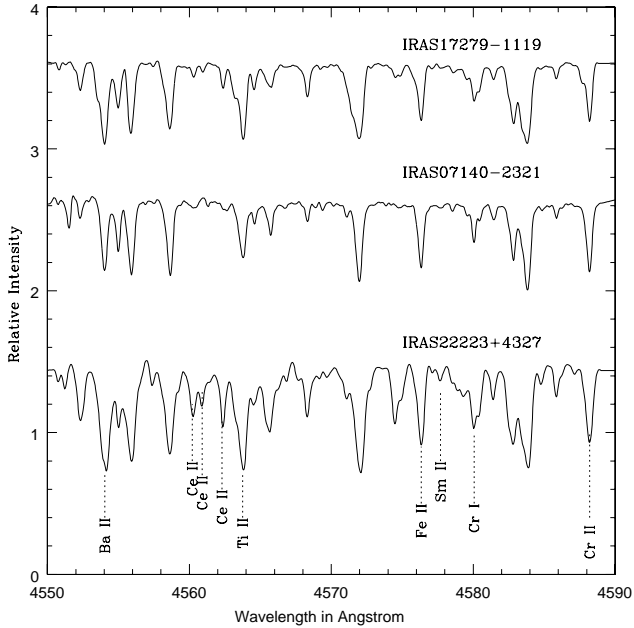}
\caption{Comparison of spectra of stars IRAS 22223+4327 and IRAS 17279-1119 showing
 s-process enrichment with IRAS 07140-2321, a normal star of similar temperature 
  in the 4550-4590 \AA\ region. The lines of s-process elements like Ba\,{\sc ii}, Ce\,{\sc ii} and Sm\,{\sc ii} are enhanced.}
\label{loci3}
\end{center}
\end{figure}

IRAS 22223+4327's progenitor was most probably a thermally pulsing AGB star. This is indicated by
the C/O ratio of
 unity and the about one dex enrichment of the $s$-process elements.
 Two of our program stars IRAS 17279-1119 and IRAS 22223+4327 show significant
 s-process enhancement. We have compared the spectra of these two objects
 with IRAS 07140-2321 with similar temperature but without s-process enhancement
in {\bf Figure \ref{loci3}.}

\subsection {BD$+39^o4926$ }

 Although this object has been mentioned in several papers on PAGB
 stars, a contemporary abundance analysis using high-quality digital
spectra, modern  model atmospheres
 and refined atomic data has not been undertaken.
Abundance data from Kodaira et al. (1970) show strong effects of
dust-gas winnowing. Yet, this star unlike other stars exhibiting
severe dust-gas winnowing does not have an infrared excess.
However, the star is a spectroscopic binary, as are many or even all
other stars exhibiting severe dust-gas winnowing. For these reasons,
 we chose to include this object in our sample.

\begin{figure}
\begin{center}
\includegraphics[width=9cm,height=9cm]{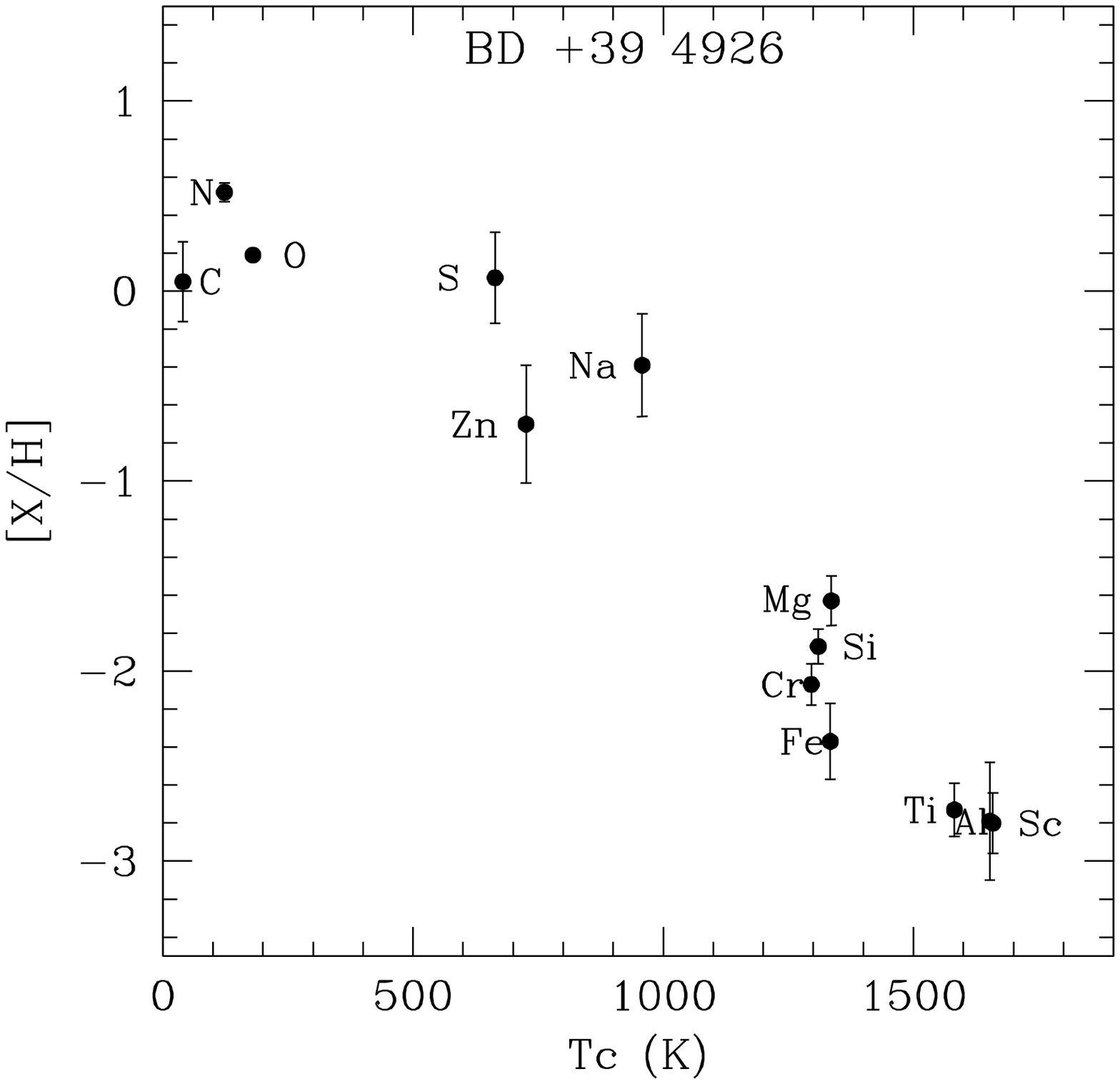}
\caption{Plot of [X/H] vs [T$_c$] for the star BD$+39^o4926$.}
\label{loci4}
\end{center}
\end{figure}

 Our spectral coverage (particularly in red region) has been more
 extensive and we could make a more detailed analysis employing
 more number of lines per species and also including the important element Zn.
 The star's final abundances are presented in Table 8.
 Figure \ref{loci4} shows the plot of [X/H] vs [T$_C$] for this star. This star shows a
 clear indication of the dust-gas separation without possessing IR
 indications of circumstellar envelope. From this plot, we infer that
 the initial metallicity of the star is $-$0.7 dex.
 (Another such star showing selective depletion
 of refractory elements without IR excess is HD~105262 (Giridhar et al. 2010)
 which has the temperature of 8000K and [Fe/H] of $-1.8$ dex.)
\subsection {[Na/Fe] in sample stars}

 The relative enrichment of Na has been observed in most of our
 program stars. The Na abundance has been measured using weak subordinate lines
 for which the estimated Non-LTE correction does not exceed
 $-$0.1 dex. The Na enrichment is believed to be caused by products of
 Ne-Na cycle involving proton capture on $^{22}$Ne in H burning region
 which are mixed to the surface following FDU. The verification of expected dependence
 of [Na/Fe] on stellar mass remains illusive.
 All our program stars give [Na/Fe] in excess of FDU predictions (e.g. El Eid and Champagne  1995 
 predict [Na/Fe] of $+$0.18 dex for 5 M$_{\odot}$ model). A further increase 
 in surface abundance of Na during AGB evolution is predicted by Mowlavi (1999)
 but the lack of correlation between the observed [Na/Fe] with s-process enhancement
 cautions against a simplistic interpretation of observed [Na/Fe]. 

\subsection {$\alpha$ elements}

 Our analysis covers whole range of elements including $\alpha$ elements which are
 important diagnostics of stellar population. Although abundances are
 estimated for all $\alpha$ elements
  Mg, Si, S, Ca and Ti, we chose to use only Mg, Si and S to compute [$\alpha$/Fe].
  The Ca and Ti abundances derived from Ca\,{\sc i}  and Ti\,{\sc i} lines could be affected by 
  Non-LTE corrections. For dwarfs and subgiants the corrections are
  estimated to be about $+$0.1 dex as described in section 4.1,
  but can become larger for supergiants. More importantly they are prone to
  depletion via grain condensation due to their large 
   condensation temperatures (T$_{C}$ for Ca and Ti are 1517 and 1582K respectively.)
 As explained in section 4.1, the
  Non-LTE correction for Mg\,{\sc i} lines used by us is about $+$0.05 dex and
  for S\,{\sc i} and Si\,{\sc i} lines it is  negligible.

 The mean [$\alpha$/Fe] values of $+$0.15 for IRAS 01259+6823 is indicative
 of thin disc population, a similar 
  value $+$0.18 for [$\alpha$/Fe] is exhibited by IRAS 04535+3747 and $+$0.13 for 
 IRAS 05208-2035 and $+$0.21 is seen for IRAS 17279-1119.  
  The [$\alpha$/Fe] of $+$0.45 for IRAS 07140-2321 is indicative of thick disc
   population; it also shows [Fe/H] of $-$0.9, 
   IRAS 07331+0021  does not have good representation of  $\alpha$ elements.
 The [$\alpha$/Fe] of $+$0.36 for IRAS 08187-1905
 and its [Fe/H] of $-$0.6  supports its
 thick disc candidature. HD~107369 also exhibits [$\alpha$/Fe] of $+$0.41 while it
 is $+$0.37 for IRAS 12538-2611 with [Fe/H] of $-$1.2. Like other s-process enhanced PAGBs 
   IRAS 22223+4327 exhibits [$\alpha$/Fe] value of $+$0.46 not withstanding its
 [Fe/H] of $-$0.3. Hence our sample stars are a mixed lot. 

\subsection{[Ca/Fe]}

For most of our sample stars as well as for a large
fraction of RV Tau stars and PAGB stars
we find [Ca/Fe] and [Ti/Fe] much lower than those expected even for thin disc stars.
 Although Fe abundance can be estimated with better precision using Fe\,{\sc ii} lines,
 for Ca, the abundance is estimated using Ca\,{\sc i} lines hence Non-LTE corrections for supergiants need to be investigated.
 (for dwarfs and subgiants $+$0.1dex is estimated but could be larger for lower gravity
  PAGB stars). Although this Non-LTE correction for supergaint gravities are not 
 available, it is comforting to note that there is no perceptible difference
 between [Ca/Fe] for a sample of giants used  by Takeda, Sato and Murata (2008) 
 and those of dwarfs and subgiants (Bensby 2005, Reddy et al. 2006). Hence we
 do not expect very large reduction in [Ca/Fe] caused by Non-LTE effect (a likely
 Non-LTE correction would be in $+$0.1 to 0.2 dex range).

Mild to severe reduction in [Ca/Fe] can be caused by the fact that the T$_{C}$ 
 for Ca and Fe are 1517K and 1334 respectively, hence Ca is relatively more
 susceptible to depletion via condensation onto the grains.

\begin{figure}
\begin{center}
\includegraphics[width=10cm,height=10cm]{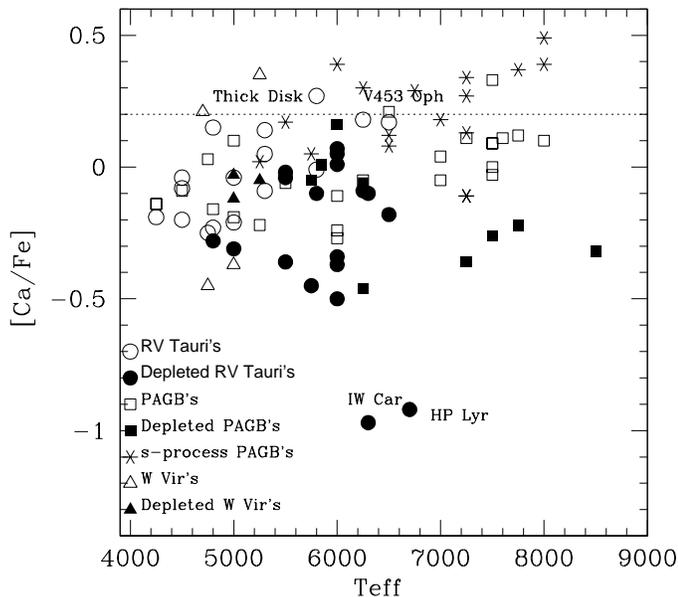}
\caption{Plot of [Ca/Fe] versus the effective temperature for all groups of PAGB's including our Program Stars.}
\label{loci5}
\end{center}
\end{figure}

To understand [Ca/Fe] variations seen in PAGBs and RV Tau stars 
a plot of [Ca/Fe] as function of temperature could be instructive notwithstanding the fact 
 that  some of the heavily depleted objects do not have
  Ca measurements and Non-LTE correction could result in
  at least $+$0.1dex vertical shift to all data points. The Figure \ref{loci5} shows
thick disc [Ca/Fe] value of $+$0.21 by dotted horizontal line.
 The post-AGB stars with s-process enhancement generally have [Ca/Fe] $>$ 0.
 For these objects,
  $\alpha$ elements including [Ca/Fe] are consistently positive. A
few PAGB stars with very mild or no s-process enrichment also show [Ca/Fe] of +0.1
to 0.0. RV Tauris such as V453 Oph showing mild s-process enhancement also
show similar [Ca/Fe] values. Depleted PAGBs and RV Tauris understandably show
negative
 [Ca/Fe]. IW Car and HP Lyr are
 well known depleted objects with very small
scatter in their T$_{C}$ vs [X/H] plots.
  But even those objects with no established indication of dust-gas
separation have small negative [Ca/Fe] in excess of what can be
 ascribed to Non-LTE effects. A large fraction of them have temperatures lower than 5000K,
 hence the signature of depletion may be muffled for them.
\section{Discussion}

 From the study of extended samples of PAGB stars it has become obvious
 that they exhibit an enormous chemical variety. However, hardcore PAGB stars
 showing the predicted outcome of the third dredge-up (increased C/O ratio and
 s-process enhancements) make a much smaller fraction of known PAGB stars.
 We have compiled the abundance data for PAGB stars and present them in three tables.
 We have tabulated separately the PAGB stars showing very distinct s-process
 enhancement ([s/Fe] $>$ 0.5 dex), those showing very distinct depletion of
 condensable elements (Depletion
 Index (DI) $>$ 1.0) and those exhibiting neither distinct s-process enhancement nor
  depletion in an attempt to statistically infer the influence of factors such as
 binarity, IR fluxes.

 \subsection{PAGB stars with s-process enhancements}

   Our compilation of  data for PAGB stars with significant s-process enhancements
   is presented in Table 9. A few stars have more than one analysis.
  Most analyses employ resolution around 40,000. More recent analyses employ more
   accurate oscillator strengths. It should be noted that a majority
  of them belong to thick disc population with [Fe/H] in range $-$0.3 to $-$0.8 dex.
  Early discoveries of this class with large [s/Fe] led to phrase "have" meaning 
  members with [s/Fe] in range $+$1.5 to $+$2.3 and "have not" with [s/Fe] $\sim$ 0.

\begin{table*}
\begin{minipage}{140mm}
\centering
\caption{A list of post-AGB stars with enhancement of s-process elements}
\label{table9}
\begin{tabular}{lllllllllll}
\hline
\multicolumn{1}{l}{IRAS} & \multicolumn{1}{l}{Other names}
& \multicolumn{1}{l}{T$_{\rm eff}$}
& \multicolumn{1}{l}{[Fe/H]} & \multicolumn{1}{l}{[$\alpha$/Fe]}
& \multicolumn{1}{l}{[s/Fe]} & \multicolumn{1}{l}{[ls/Fe]} &
 \multicolumn{1}{l}{[hs/Fe]}
& \multicolumn{1}{l}{[hs/ls]} & \multicolumn{1}{l}{Ref}
& \multicolumn{1}{l}{Binarity} \\
\hline
08281-4850& ...& 7750& $-$0.33& $+$0.25&
 $+$1.24& $+$1.23& $+$1.25& $+$0.02& 8& No \\
22223+4327& BD+42$^{o}$4388, V448 Lac& 6500&
 $-$0.33& $+$0.25& $+$0.92& $+$1.31& $+$1.18& $-$0.13& 1& No \\
...& ...& 6750& $-$0.30& $+$0.24& $+$1.08&
 $+$1.36& $+$0.94& $-$0.42& 9 & No \\
08143-4406& ...& 7250& $-$0.39& $+$0.24&
 $+$1.51& $+$1.53& $+$1.50& $-$0.02& 2& No \\
06530-0213& ...& 7250& $-$0.46& $+$0.20&
 $+$2.06& $+$1.83& $+$2.18& $+$0.35& 2& No \\
17279-1119& HD 158616, V340 Ser& 7250& $-$0.43&
 $+$0.25& $+$0.64& $+$0.70& $+$0.62& $-$0.08& 1& No \\
...& ...& 7300& $-$0.60& $+$0.44& $+$0.69&
 $+$0.95& $+$0.60& $-$0.35& 5& No \\
...& ...& 7400& $-$0.68& $+$0.78& $+$0.50&
 $+$0.50& ..& $-$0.50& 6& No \\
Z02229+6208& ...& 5500& $-$0.50& $+$0.17&
 $+$1.37& $+$2.08& $+$1.13& $-$0.95& 4& No \\
07430+1115& ...& 6000& $-$0.50& $+$0.09& $+$1.60&
 $+$2.09& $+$1.44& $-$0.65& 4& No \\
14325-6428& ...& 8000& $-$0.55& $+$0.30& $+$1.25& $+$1.23& $+$1.29
& $+$0.06& 8& No \\
18384-2800& HD 172481, V4728 Sgr& 7250& $-$0.62&
 $+$0.46& $+$0.48& $+$0.58& $+$0.44& $-$0.14& 5& Yes \\
...& ...& 7250& $-$0.55& $+$0.37& $+$0.48&
 $+$0.49& $+$0.47& $-$0.02& 9& Yes \\
04296+3429 & ...& 7000& $-$0.65& $+$0.38& $+$1.55& $+$1.66&
 $+$1.49& $-$0.17& 9 & No \\
19500-1709& HD 187885, V5112 Sgr& 8000& $-$0.66&
 $+$0.30& $+$1.15& $+$1.40& $+$1.03& $-$0.37& 9 & No \\
...& ...& 8000& $-$0.44& $+$0.54& $+$1.31&
 $+$1.29& $+$1.35& $+$0.06& 7& No \\
05341+0852& ...& 6500& $-$0.72& $+$0.29& $+$2.22&
 $+$2.01& $+$2.32& $+$0.32& 9 & No \\
05113+1347& ...& 5250& $-$0.75& $+$0.18& $+$2.00&
 $+$1.61& $+$2.16& $+$0.55& 3& No \\
23304+6147& ...& 6750& $-$0.81& $+$0.32& $+$1.60&
 $+$1.55& $+$1.63& $+$0.09& 9 & No \\
22272+5435& HD 235858, V354 Lac& 5750& $-$0.82&
 $+$0.17& $+$1.99& $+$1.56& $+$2.17& $+$0.61& 3& No \\
07134+1005& HD 56126, CY CMi& 7250& $-$1.03&
 $+$0.08& $+$1.51& $+$1.56& $+$1.49& $-$0.06& 9 & No\\
\hline
\end{tabular}
\flushleft$^{1}${Present work.}, $^{2}${Reyniers et al. 2004}, $^{3}${Reddy et al. 2002}, $^{4}${Reddy et al. 1999}
\flushleft$^{5}${Arellano Ferro et al. 2001}, $^{6}${Van Winckel et al. 1997}, $^{7}${Van Winckel et al. 1996}
\flushleft$^{8}${Reyniers et al. 2007}, $^{9}${Reyniers (2002)}
\end{minipage}
\end{table*}

 However, recent studies  (including the present work) have found the objects with
  moderate s-process enhancements and C/O not exceeding one. Heavily enriched objects
 seem to favour a temperature range of 6000K to 7500K. They all have [$\alpha$/Fe] in
  similar to those of thick disc stars.
  A large fraction of them show 21$\mu$ feature.
  The SED generally contains well resolved IR component of strength comparable to
  photospheric component. Three objects from Table 9 IRAS 08143-4406, IRAS 22223+4327
  and IRAS 23104+6147 have nebulae of SOLE class (generally of 2 arcsec size)
  detected by Si\'{o}dmiak et al. (2008).
                           Another interesting feature of this class of s-process
  enhanced PAGB stars is that very few of them are known binaries. They appear to
 represent single star evolution of moderately metal-poor thick disc stars.
 \subsection{PAGB stars with depletion of refractory elements}

  Although a working hypothesis of dust-gas separation in circumbinary disc
  causing selective depletion of refractory elements has been proposed,
  the actual mechanism of disc formation and its evolution
  is far from understood. 

  We have compiled the existing
  data  for all known PAGB stars and RV Tau stars and 
  chosen very stringent selection criteria to define the group of
  PAGB stars showing depletion due to selective removal of condensable elements.
 The guideline being [S/Fe] and [Zn/Fe] $>$ 0 and [Ca/Fe], [Sc/Fe],
[Ti/Fe] $<$ 0.0.  
 We present our compilation in Table 10.
  It is possible that adherence to this criteria would result in non- inclusion
  of some PAGB stars considered mildly depleted in other studies. 

 Quantifying the depletion process is a very difficult task. S and Zn are good representatives
 of non-depleted elements due to their low condensation temperatures
 704 and 726K respectively. CNO inspite their lower T$_{C}$ are not useful since they are 
 affected by nuclear processing and dredge-ups. However Zn\,{\sc i} lines are strongest at
 around 4700K and remain strong at even lower temperatures but the spectrum gets 
 too crowded to measure them. At higher temperatures like 7000K and above they become
 weak and are weaker than 10 m\AA\ at 8000K. The lines of S\,{\sc i} are at the strongest near 6700K
 but weaken drastically at cooler as well as hotter temperatures. Hence the abundance errors are
 highly temperature dependent.
 We have taken mean of heavily depleted elements Ca, Sc and Ti
 (with similar condensation temperatures 1517, 1582 and 1659K respectively) and subtracted it from
 [S/H] to define depletion index given in Table 10.
 Ideally [Al/H] should have been included in the calculations, but Al abundances
 were not available for many stars. For a few stars only two of these three elements
 are  measured; then the mean is taken of the elements for which data is available.
 For heavily depleted stars [Zn/H] is a better metallicity indicator; hence the objects with
  [Zn/H] $<$ $-$0.5 dex could belong to thick disc and halo population.

 We have studied T$_{C}$ vs [X/H] curves for all stars of Table 10 and have
  indicated by $*$ those that exhibit very smooth curves like HP Lyr
  in the column for DI (Depletion Index).

 The compilation brings about intriguing statistics; out of 36 depleted PAGB stars
 25 are known binaries,
 28 have IRAS detections seven are without (which is about 19\%).

 Hence it is obvious that binarity plays very important role. Non detection
 of IRAS fluxes could be caused by weakening and cooling of the disc
 hence surveys at longer wavelengths and with increased sensitivities are
 required to resolve this issue.

 However there are objects like 
  IRAS 07140-2321 which has dusty CS, known to be
 a binary with period of 116 days and yet does not show the depletion!
 The boundary conditions suggested in scenarios proposed to explain these 
 phenomenon deserve a close scrutiny.

\begin{table*}
\begin{minipage}{170mm}
\centering
\caption{A list of post-AGB stars showing depletion}
\label{table10}
\begin{tabular}{llllllllllllll}
\hline
\multicolumn{1}{l}{IRAS}&
\multicolumn{1}{l}{Other$^{a}$}&
\multicolumn{1}{l}{T$_{\rm eff}$}&
\multicolumn{1}{l}{[12]-[25]}&
\multicolumn{1}{l}{[25]-[60]}&
\multicolumn{1}{l}{[S/H]}&
\multicolumn{1}{l}{[Zn/H]}&
\multicolumn{1}{l}{[Ca/H]}&
\multicolumn{1}{l}{[Sc/H]}&
\multicolumn{1}{l}{[Ti/H]}&
\multicolumn{1}{l}{[Fe/H]}&
\multicolumn{1}{l}{DI$^{b}$}&
\multicolumn{1}{l}{Ref}&
\multicolumn{1}{l}{Bin$^{c}$} \\
\hline
...& TT Oph& 4800& ..& ..& $+$0.01& $-$0.71& $-$1.13 & $-$1.09& $-$0.82& $-$0.85& 1.02 &2  & N \\
...& UZ Oph& 5000& ..& ..& $-$0.38&$-$0.74&$-$1.10&$-$1.26&$-$1.00&$-$0.79& 1.50& 4& N \\
...& W Vir& 5000& ..& ..& $-$0.26&$-$0.84&$-$1.03&$-$2.03&$-$1.08&$-$1.00& 1.12& 12& N \\
...& V1711 Sgr& 5000& ..& ..& $-$0.07& $-$0.98& $-$1.32& $-$2.46&$-$2.05& $-$1.20& 1.87& 12& N \\
...& RX Lib& 5250& ..& ..& $-$0.58& $-$0.80&$-$1.05&$-$2.22&$-$1.14&$-$1.00& 0.89& 12& N \\
17250-5951& UY Ara& 5500& $+$0.27& $-$0.95& $+$0.01&
 $-$0.29& $-$1.06& $-$1.74& ..& $-$1.02&1.41$^{\ast}$ &2 & N   \\
06054+2237& SS Gem& 5500& $-$0.28&
$+$0.08& $-$0.42& $+$0.02& $-$1.05& $-$1.92& $-$2.00& $-$0.87& 1.24 &3 & N  \\
06072+0953& CT Ori& 5500& $-$0.11& $-$1.63&
 $-$0.53& $-$0.58& $-$1.80& $-$2.58& $-$2.51& $-$1.87& 1.65$^{\ast}$ &3 & Y  \\
06160-1701& UY CMa& 5500& $-$0.38& $-$1.59&
 $-$0.32& $-$0.59& $-$1.65& $-$2.22& $-$2.38& $-$1.29&1.65$^{\ast}$ &4  & Y  \\
20117+1634& R Sge& 5750& $-$0.37& $-$1.38& $+$0.37&$-$0.19&$-$0.95&$-$1.48&$-$1.34&$-$0.50&1.63& 5& Y \\
19163+2745& EP Lyr& 5750& $+$0.52& $-$0.24&
 $-$0.61& $-$0.70& $-$1.82& $-$2.11& $-$2.01& $-$1.80&1.26$^{\ast}$ & 5  & Y \\
09144-4933& ..& 5750& $-$0.25& $-$1.57&$-$0.01& ...& $-$0.37&$-$1.65& $-$1.28& $-$0.31& 1.09 &14 & N \\
20056+1834& QY Sge& 5850& $+$0.03& $-$1.31& $+$0.14& $-$0.14& $-$0.28& $-$0.70& $-$1.15& $-$0.29&0.85 & 13 & Y \\
07008+1050& HD 52961& 6000& $-$0.77&$-$0.92& $-$1.00&$-$1.40& ..& ..& ..& $-$4.80&3.80$\dagger$ &8 & Y \\
06338+5333& HD 46703& 6000& $-$0.21& $+$0.05& $-$0.40&
 $-$1.40& $-$1.60& ..& $-$1.79& $-$1.60& 1.00$^{\ast}$ &11& Y \\
18281+2149& AC Her & 6000& $+$0.50& $-$1.21&
 $-$0.37& $-$0.93& $-$1.50& $-$1.70& $-$1.64& $-$1.40&1.13$^{\ast}$ &6 & Y \\
08011-3627& AR Pup& 6000& $-$0.36& $-$1.39&
 $+$0.44& ..& $-$1.37& $-$2.16& ..& $-$0.87&2.20 & 5  & Y \\
06034+1354& DY Ori& 6000& $+$0.19& $-$1.38&
 $+$0.16& $+$0.21& $-$1.70& ..& $-$2.33& $-$2.30&2.65$^{\ast}$ & 4  & Y \\
12067-4508& RU Cen, & 6000& $+$0.78&
 $-$0.72& $-$0.68& $-$1.00& $-$1.89& $-$1.88& $-$1.96& $-$1.90&1.11 &7   & Y \\
18548-0552& BZ Sct& 6250& $+$0.05& $+$0.49& $+$0.18&$+$0.04&$-$0.91&$-$1.13&$-$1.22& $-$0.82&1.27& 4& N \\ 
   ...& CC Lyr& 6250& ..& ..& $-$0.50&
 $-$1.20& ..& ..& ..& $-$3.40& 2.90$\dagger$& 12 & N \\
16230-3410& ...& 6250& $-$0.27&$-$1.52& $-$0.36&$-$0.42& $-$0.74& $-$2.28& $-$1.45& $-$0.68&1.13 &14 & Y \\
17233-4330& ...& 6250& $-$0.26& $-$1.40& $+$0.14& $-$0.21& $-$1.39& $-$1.60& $-$1.62& $-$0.98& 1.68&14 & Y \\
17243-4348& LR Sco& 6250& $-$0.22& $-$0.94& $+$0.04& $+$0.22& $-$0.17& $-$1.11& $-$0.61& $-$0.05&1.07 & 4 & Y \\
18564-0814& AD Aql& 6300& $+$0.09& $-$1.55&
 $-$0.03& $-$0.12& $-$2.22& $-$1.79& $-$2.57& $-$2.12&2.16$^{\ast}$ & 6  & Y \\
19199+3950& HP Lyr& 6300& $-$0.02& $-$1.56&
 $+$0.05& $-$0.35& $-$1.95& $-$2.87& $-$2.97& $-$0.98&2.83$^{\ast}$ &4   & N \\
12185-4856& SX Cen & 6500& $-$0.54&
 $-$1.28& $-$0.08& $-$0.54& $-$1.52& $-$1.96& $-$1.97& $-$1.15&1.62 & 7  & Y \\
09256-6324& IW Car, & 6700& $-$0.05&
 $-$1.11& $+$0.36& $-$0.04& $-$1.97& $-$2.13& ..& $-$1.0&2.41$^{\ast}$ & 10 & Y \\
08544-4431& V390 Vel& 7250& $-$0.14& $-$1.13&
 $+$0.12& $+$0.09& $-$0.36& $-$0.87& $-$0.85& $-$0.50&0.81$^{\ast}$ &15 & Y \\
15469-5311& ..& 7500& $-$0.16& $-$1.08& $+$0.51& $+$0.25&$-$0.46&$-$1.56&$-$1.57&$+$0.04& 1.70& 14& Y \\
10158-2844& HR 4049& 7500& $-$1.76& $-$1.83&
$-$0.40& $-$1.30& $-$5.30& ..& ..& $-$4.80&4.90$^{\ast}$ &8 & Y \\
06176-1036& HD 44179& 7500& $+$0.09& $-$1.05&
 $-$0.30& $-$0.60& $-$3.11& ..& ..& $-$3.30&2.69$^{\ast}$ &8& Y \\
19125+0343& BD+03$^{o}$3950& 7750& $-$0.09&
 $-$1.32& $+$0.48& $+$0.13& $-$0.51& ..& $-$2.15& $-$0.35&1.75$^{\ast}$ &14& Y \\
  ...&  BD+39$^{o}$4926& 7750& ..& ..& $+$0.07&
 $-$0.70& ..& $-$2.80& $-$2.73& $-$2.37&2.72$^{\ast}$ & 1 & Y \\
22327-1731& HM Aqr & 8200& $-$0.19&
 $-$0.86& $+$0.36& ..& ..& $-$1.31& $-$1.49& $-$0.90&1.76 &8& Y \\
   ...& HD 105262& 8500& ..& ..& $-$0.50& ..&
 $-$2.10& $-$2.07& $-$1.58& $-$1.90&1.42$^{\ast}$ &9& Y \\
\hline
\end{tabular}
\flushleft$^{a}${Other refers to other names.}
\flushleft$^{b}${DI refers to Depletion Index. $\dagger$ refers to D.I measured from S and Fe in absence of other elements. $\ast$ indicate that the depletion plots of these stars had less scatter.}
\flushleft$^{c}${Bin refers to Binarity.}
\flushleft$^{1}${Present work.}, $^{2}${Giridhar et al. 2000}, $^{3}${Guillermo Gonzalez et al. 1997a}, $^{4}${Giridhar et al. 2005}
\flushleft$^{5}${Guillermo Gonzalez et al. 1997b}, $^{6}${Giridhar et al. 1998}, $^{7}${Maas et al. 2002}, $^{8}${Van Winckel (1995)}
\flushleft$^{9}${Giridhar et al. 2010}, $^{10}${Giridhar et al. 1994}, $^{11}${Hrivnak et al. 2008}, 
\flushleft$^{10}${Giridhar et al. 1994}, $^{11}${Hrivnak et al. 2008}, $^{12}${Maas et al. 2007}
\flushleft$^{13}${Rao et al. 2002}, $^{14}${Maas et al. 2005}, $^{15}${Maas et al. 2003}

\end{minipage}
\end{table*}

\begin{table*}
\begin{minipage}{160mm}
\centering
\caption{A list of post-AGB stars showing neither significant s-process enrichment 
 nor  depletion }
\label{table11}
\begin{tabular}{lllllllllllll}
\hline
\multicolumn{1}{l}{IRAS}&
\multicolumn{1}{l}{Other$^{a}$}&
\multicolumn{1}{l}{T$_{\rm eff}$}&
\multicolumn{1}{l}{[12]-[25]}&
\multicolumn{1}{l}{[25]-[60]}&
\multicolumn{1}{l}{[S/H]}&
\multicolumn{1}{l}{[Zn/H]}&
\multicolumn{1}{l}{[Ca/H]}&
\multicolumn{1}{l}{[Sc/H]}&
\multicolumn{1}{l}{[Ti/H]}&
\multicolumn{1}{l}{[Fe/H]}&
\multicolumn{1}{l}{Ref}&
\multicolumn{1}{l}{Bin$^{b}$} \\
\hline
02008+4205& HD 12533& 4250& $-$1.55& $-$2.05& ..& $-$0.05&$-$0.20&$-$0.08&$-$0.21& $-$0.06& 8& N\\
19437-1104& DY Aql& 4250& $+$0.17&$-$0.37&..& ..& $-$1.22& $-$2.13& ..& $-$1.03& 3& N \\
20343+2625& V Vul& 4500& $-$0.84&$-$1.60& $+$0.57&$-$0.27&$-$0.47&$-$0.69&$-$0.18&$-$0.39& 1& Y \\
18448-0545& R Sct& 4500& $-$0.89& $-$0.14& ..& $-$0.19& ..& $-$1.43&$-$0.38&$-$0.35& 2& N \\
04440+2605& RV Tau& 4500& $-$0.24&$-$1.11& ..& $+$0.02&$-$0.48&$-$0.35&$-$0.53&$-$0.40& 2& Y \\
07331+0021& AI CMi& 4500& $+$1.62& $-$1.42& ..& ..& $-$1.25& $-$1.35& $-$0.89& $-$1.16& 11& N \\
06489-0118& SZ Mon& 4700& $-$0.54& $-$1.38& $+$0.24& $-$0.43& $-$0.19& $-$1.46& $-$1.25& $-$0.40& 10& N \\
...& MZ Cyg& 4750& ..& ..& $+$0.53& $-$0.23& $-$0.65& $-$1.07& $-$0.63& $-$0.20&10& N \\
...& AZ Sgr& 4750& ..& ..& $-$0.34&..& $-$1.80&$-$1.78&$-$1.57&$-$1.55& 1& N \\
17038-4815& ..& 4750& $+$0.19&$-$0.84& ..& $-$1.20&$-$1.47&$-$2.00&$-$1.40&$-$1.50& 6& Y\\
17193+8439& HD 159251& 4800& $-$1.27&$+$0.99& ..& $-$0.14& $-$0.21& $-$0.03& $-$0.10& $-$0.05& 8& N \\
19472+4254& DF Cyg& 4800&$-$0.65&$-$0.82&..& $-$0.62&$-$0.23&$-$0.96&$-$0.04&$+$0.00& 1& N  \\
04166+5719& TW Cam& 4800& $-$0.42&$-$1.20& $-$0.05&$-$0.34&$-$0.65&$-$0.43&$-$0.64& $-$0.50& 2& Y \\
07284-0940& U Mon& 5000& $-$0.37&$-$1.31& $-$0.15&$-$0.71&$-$0.96&$-$0.97&$-$0.70&$-$0.79&2& Y \\
F17015+0503& TX Oph& 5000& ..& ..& $-$0.63&$-$1.23&$-$1.43&$-$1.79&$-$1.18&$-$1.22& 1&N  \\
01369+4124& HD 10132& 5000& $-$0.62& $+$0.02& $+$0.15& $-$0.14& $+$0.05& $+$0.36&$-$0.02&$-$0.05& 8& N \\
...& CO Pup& 5000& ..& ..& $-$0.05& $-$0.73& $-$0.97& $-$1.56& $-$0.94& $-$0.60&10& N \\
...& TW Cap& 5250& ..& ..& $-$1.40& $-$1.52& $-$1.45& $-$1.61& $-$1.42& $-$1.80&10& N \\
...& V360 Cyg& 5250& ..& ..& $-$0.88& $-$1.36&$-$1.26& ..& $-$1.28& $-$1.40& 4& N \\
17530-3348& AI Sco& 5300& $-$0.47&$-$1.46&$-$0.08&$-$0.61&$-$0.64&$-$0.96&$-$0.89&$-$0.69& 1& Y \\
...& AR Sgr& 5300& ..& ..& $-$0.82&$-$1.20&$-$1.44&$-$1.41&$-$1.21&$-$1.33& 1& N \\
09538-7622& ...& 5500& $-$0.40& $-$1.20& $-$0.30&$-$0.60&$-$0.66& $-$0.90&$-$0.70& $-$0.60& 6& Y \\
...& RX Cap& 5800& ..& ..& $-$0.57&$-$0.63&$-$0.79&$-$1.20&$-$0.62&$-$0.78& 1& N  \\
...& BT Lib& 5800& ..& ..& $-$0.76&$-$1.06&$-$0.91&$-$0.71&$-$0.96&$-$1.18& 2& N \\
04535+3747& V409 Aur& 6000& $-$0.73& $-$0.03& $-$0.16& $-$0.53& $-$0.59& $-$0.75& $-$0.63& $-$0.48& 11& N \\
13110-5425& HD 114855& 6000& $+$2.08&$+$1.39&$+$0.34&$-$0.26&$-$0.35&$-$0.34&$-$0.41&$-$0.11& 5& N \\
14524-6838& EN TrA& 6000& $-$0.28&$-$0.99&$-$0.61&$-$0.56&$-$1.03&..& $-$1.11&$-$0.76& 7& Y\\
08187-1905& V552 Pup& 6250&$+$ 3.49& $-$0.39& $+$0.05& $-$0.58& $-$0.63& $-$0.75& $-$0.99& $-$0.59& 11& N \\
09060-2807& BZ Pyx& 6500& $-$0.43& $-$1.46& $-$0.70& $-$0.50& $-$0.49&$-$0.80&$-$0.70&$-$0.70& 6& Y \\
...& DS Aqr& 6500& ..& ..& $-$0.82& $-$1.07&$-$0.97&$-$1.17&..&$-$1.14& 2& N \\
12222-4652& HD 108015& 7000& $+$0.03& $-$1.55&$-$0.15&$-$0.16&$-$0.05&$-$0.80&$-$0.26&$-$0.09& 7& Y \\
IRAS 07140-2321& SAO 173329& 7000& $-$0.24& $-$1.14& $-$0.32& $-$0.87& $-$0.97& $-$0.92& $-$0.73& $-$0.92& 11& Y \\
04002+5901& HD 25291& 7250& $-$1.46&$-$0.02& $+$0.08& ..& $-$0.22& $+$0.01& $-$0.30& $-$0.33& 8& N \\
07018-0513& HD 53300& 7500& $-$0.97&$+$0.30&$-$0.04&$-$0.71&$-$0.62&$-$0.96&$-$0.83&$-$0.62& 5& N \\
18439-1010& HD 173638& 7500& $-$1.38& $+$0.59& ..& $-$0.06&$+$0.01& $+$0.16&$-$0.05& $-$0.08& 9& N\\ 
12175-5338& SAO 239853& 7500& $+$3.25& $-$1.08&$-$0.40&$-$0.73&$-$0.48&$-$1.11&$-$0.54& $-$0.81 & 7& N  \\
04175+3827& HD 27381& 7500& $-$0.99& $+$0.24& $-$0.30& ..& $-$0.59& $-$0.39&$-$0.57&$-$0.67& 8& N \\
...& HD 107369& 7500& ..& ..& $-$0.84& ..& $-$1.36& $-$1.45& $-$1.09& $-$1.33& 11& N \\
11000-6153& HD 95767& 7600& $-$0.37&$-$0.39&$+$0.01&$-$0.12&$+$0.24&..&$-$0.11&$+$0.13& 7& Y \\
19157-0247& ...& 7750& $-$0.23& $-$1.17& $+$0.20& ..& ..& $-$0.40&$-$0.20&$+$0.10& 6& Y \\
...& HD 10285& 7750& ..& ..& $+$0.04&$-$0.36&$-$0.19&$-$0.07&$-$0.30&$-$0.31& 8& N \\
...& HD 218753& 8000& ..& ..& $+$0.17&$+$0.18& $-$0.09& $+$0.00& $-$0.26&$-$0.19& 9& N \\
\hline
\end{tabular}
\flushleft$^{a}${Other refers to other names}, $^{b}${Bin refers to Binarity.}
\flushleft$^{1}${Giridhar et al. 2005}, $^{2}${Giridhar et al. 2000}, $^{3}${Guillermo Gonzalez et al. 1997a}, $^{4}${Giridhar et al. 1998}
\flushleft$^{5}${Giridhar et al. 2010}, $^{6}${Maas et al. 2005}, $^{7}${Van Winckel (1996)}, $^{8}${Giridhar \& Arellano Ferro (2005)}
\flushleft$^{9}${Arellano Ferro et al. 2001}, $^{10}${Maas et al. 2007}, $^{11}${Present Work}

\end{minipage}
\end{table*}

 In our study of extended sample for RV Tau stars (Giridhar et al. 2005) we had proposed
 two scenarios (based upon the single star and binary configuration)
 and also discussed the parameters strongly affecting the depletion
 caused by dust-gas separation. 

 \begin{description}

  \item {$\bullet$} {\bf The surface temperature:} 
 The first important requirement was that the
 star must be hotter than $\sim$ 5000K.
  The cooler RV Tau stars (RVA) did not show the
 abundance anomalies,
  possibly their massive convective envelopes diluted the effect of accreted gas which
 was cleaned of refractory elements. The masses of convective envelopes for PAGB stars
 increases with decreased temperature (see Frankowski 2003 for more details) varies from
0.016 M$_{\odot}$ at 4000K to 0.0001  M$_{\odot}$ at 6000K. At 6000K the mass of convective
 envelope is about 1000 times that of observable photosphere. At hot temperature the
 abundance deficiency of readily condensable elements (those with high T$_{c}$) may
 approach a thousandfold and in such cases the surface and convective envelope would
be composed of almost undiluted accreted gas. At lower temperatures, in addition to
 larger convective mass the stronger stellar wind may oppose the accretion of the clean gas.

 \item {$\bullet$}  {\bf The metallicity:}
 The dust-gas separation process has not been seen in stars with intrinsic metallicity
 lower than $\sim$ $-$1.0 irrespective of their temperatures. It is likely consequence
 of the inability of radiation pressure on dust grains to force a separation of dust
 from gas when the mass fraction of the dust is very low as seen in metal-poor environments.
 A dust-gas reservoir with metallicity in excess of $-$1.0 is mandatory for efficient
 separation of dust from the gas. These conditions can be easily realized in binary scenario
 with circumbinary disc since it may contain gas from either binary companion or the main star
before the onset of accretion.

  \end {description}

Although  a revised  minimum temperature 4800K for discernible depletion  has been
identified, but stars in temperature range 6000K to 7500K tend to
 show large depletions. At hotter side {\itshape T$_{eff}$}$>$ 8500K DI is smaller; a shift
 in ionization structure causing  the observable
features to  become too scarce for S and Zn may be responsible.

Many stars such as CC Lyr, HR 4049, BD+39$^{o}$4926 show large differences
 between S and Zn despite these elements having nearly the same T$_{C}$.
 For thick disc and halo stars relative enrichment of S by $+$0.3 dex can be anticipated
  since the S is an $\alpha$ element but for larger differences found in the above
 mentioned objects there is no ready explanation.
 
 Another intriguing observation is that DI are not correlated with IRAS colours
 nor do they show any correlation with IR flux at a given wavelength.

 Since the SED characteristics can be used to develop a model for circumstellar material,
 De Ruyter et al. (2006) have done optical photometry and have also used the IR and
 sub-millimeter data for a large sample of known binary PAGB stars and other objects
 (like RV Tau stars) of similar IR characteristics. Using the optically thin dust model
they found the dust at or near sublimation temperature (1200K)
 and also very close to the star ($\sim$ 10 AU from the central source) irrespective
 of the effective temperature of the star.  These authors argued that at least a
 part of the dust must be gravitationally bound since any typical AGB outflow velocity
 would bring the dust to cooler region within years. Although their sample of known
 PAGB binary and RV Tau stars of similar characteristics had a large range in the
 size of IR excess, but the shape of IR excess indicated that CS was not freely expanding
 but was stored in the system. A study of orbital parameters for the well-studied objects
 point to orbits which are too compact for sizes of PAGB stars. These authors further
 suggest that it is highly improbable that these stars have evolved as single stars;
 a strong interaction phase while the star was in giant stage is required to explain the
 present configuration.

 The high spatial resolution interferometry in mid IR has been used to resolve compact
 dusty discs around the depleted binary post-AGB stars. For Red Rectangle, HR 4049
 and recently for HD 52961 the dusty discs have been resolved (Deroo et al. 2007).
 It is comforting to see the additional information provided by high
 spatial resolution interferometry towards better understanding of these systems
but the accuracies of disc dimensions are constrained by the radiative 
 transfer models.

\subsection{PAGB stars without significant s-process enhancement nor showing
 depletions}

 A good number of PAGB candidates displaying  basic PAGB characteristics
 like  IRAS colours, high luminosity, small amplitude light and radial 
 velocity variations do not belong to the above mentioned classes. Regarding 
  s-process enhancements, the
  earlier samples of PAGB belonged to only two classes "have" and "have not";
 the "have" showing strong s-process enhancements ([s/Fe] $>$ 1] while many
 of the "have nots" had subsolar [s/Fe] ratios.  With increased sample
 the moderately enhanced members are making an appearance. IRAS 17279-1119 has been
 known for some time but IRAS 01259+6823, IRAS 05208-2035 and IRAS 08187-1905 are new additions.

 Then there are other PAGB stars including RV Tauris without significant
  s-process enhancement nor showing depletions. We have tabulated 42 such
 objects (more than half are RV Tauris and  Semi-regular variable of type D  SRDs)
  in Table 11. A large fraction of them (30) are IRAS sources,
  One fourth (13) are binaries all showing IRAS fluxes.
  In terms of temperature range and IRAS colours they are indistinguishable
 from those in Table 10. PAGB with  s-process enhancements (Table 9)
 appear to be systematically warmer. It should be noted that larger 
 DI PAGB also seem to favour the same temperature range. Then it would 
 appear that condition of low surface temperature are not conducive
 for the survival of abundance peculiarities although unglamorous objects
 are found even in among objects in temperatures range of 6000K to 7500K.

\section{Summary}

  In our exploration of post-AGB candidates we find
   mild enhancements of s-process elements
   in IRAS 01259+6823, IRAS 05208-2035 and IRAS 08187-1905, while significant  enhancement is
   found for IRAS 17279-1119 and IRAS 22223+4327.
    Mild s-process enhancement and low temperatures of IRAS 05208-2035 and IRAS 01259+6823
    are not of common occurrence in post-AGB stars although the theory does not prohibit such objects
    since the degree of s-process enhancement gradually builds with number of
    thermal pulses. However further evidence like intrinsic
     luminosity estimates are awaited.

   A survey of well established s-process enhanced post-AGB stars shows majority of this group
    to be most likely thick disc objects of intermediate temperatures.
   It does not appear that their evolution is strongly affected by the presence
   of binary companion. Most of these
    heavily s-processed objects have temperatures between 6000K to 7500K,
    although a spread in [s/Fe] for similar atmospheric parameters is seen
    indicating the influence of additional processes on third dredge-up efficiency.

A contemporary analysis of BD+39$^{o}$ 4926 has been conducted and initial metallicity of $-$0.7 dex has been estimated. 
 HD 107369 appear to
  be a metal-poor, high galactic latitude supergiant with strong influence of CN processing.
  The observed N enhancement is in excess of that predicted by the FDU 
  but cannot be ascribed to large rotation at main sequence as the stellar lines are very
  sharp and narrow $-$ not much broader than the instrumental profile.

   Our compilation of the PAGB stars showing the abundance peculiarities caused
  by selective depletion of refractory elements reiterates the temperature and metallicity
  limits observed earlier for their detection.  The
   observed spectral energy distribution with dust excess of these
   objects have been interpreted by De Ruyter et al. 2006 via a
   circumstellar dust shell contained in Keplerian rotating discs. But the actual
   resolution of dusty discs and measurement of their sizes has been possible only
   for heavily depleted objects like HR 4049, Red Rectangle and more recently HD 52961
   although HD 52961 does not exhibit large IR excess. Search of such depleted objects
    and their follow-up using
    high spatial resolution mid IR interferometry and IR spectrometry can
    provide  valuable insight into the circumstellar geometries of these objects.

    Our compilation of PAGB stars showing neither of these two peculiarities is
    indistinguishable from these two samples in terms of basic stellar parameters.
     Hence  the cause separating these subgroups among PAGB stars is not understood.

    The extension of abundance analysis to PAGB candidates located in uniform stellar system
    such as globular clusters, or in neighbouring galaxies would help 
    since each system would have small range in mass and metallicity.  

\section*{Acknowledgments}
 DLL thanks the Robert A. Welch Foundation of Houston, Texas for support
through grant F-634.



\begin{thebibliography}{99}
\bibitem[]{} Aldenius, M., Lundberg, H., Blackwell-Whitehead, R., 2009, A\&A,
 502, 989
\bibitem[]{} Allen, D. M., Porto de Mello, G. F., 2011, A\&A, 525, 63
\bibitem[]{} Arellano Ferro, A., Giridhar, S., Mathias, P., 2001, A\&A, 368, 250
\bibitem[]{} Arkhipova, V. P., Noskova, R. I., Ikonnikova, N. P., Komissarova, G. V., 2003, Astronomy Letters, 29, 480
\bibitem[]{} Asplund, M., Grevesse, N., Sauval, A. J., 2005, ASP Conf.Ser.336, 25
\bibitem[]{} Bensby, T., Feltzing, S., Lundstr\"{o}m, I., Ilyin, I., 2005, A\&A, 433, 185
\bibitem[]{} Bergemann, M., Cescutti, G., 2010, A\&A, 522, A9
\bibitem[]{} Bergemann, M., 2011, MNRAS, 413, 2184
\bibitem[]{} Bi\'{e}mont, E., Grevesse, N., Hannaford, P., Lowe, R. M.,
1981, ApJ, 248, 867
\bibitem[]{} Bogaert, E., Ph.D. Thesis, K.U. Leuven, 1994
\bibitem[]{} Boyarchuk, A. A., Lyubimkov, L. S., Sakhibullin, N. A., 1985, Astrophysics, 22, 203
\bibitem[]{} Bl\"{o}cker, T., Sch\"{o}nberner, D., 1991, A\&A, 244, L43 
\bibitem[]{} Curry, J.J., 2004, J. Phy. Chem. Ref. Data,  33, 725
\bibitem[]{} Den Hartog, E. A., Lawler, J. E., Sneden, C., Cowan, J. J,
 2003, ApJS, 148, 543
\bibitem[]{} Deroo, P., Ache, B., Verhoelst, T., Dominik, C., Tatulli, E., Van Winckel, H., 2007, A\&A, 474, 45
\bibitem[]{} De Ruyter, S., Van Winckel, H., Maas, T., Llyod Evans, T., Waters, L. B. F. M., Dejonghe, H.,  
 2006, A\&A, 448, 641
\bibitem[]{} El Eid, M. F., Champagne, A. E., 1995, ApJ, 451, 298
\bibitem[]{} Frankowski, A., 2003, A\&A, 406, 265
\bibitem[]{} Fuhr, J. R., Martin, G. A., Wiese, W. L., 1988, J. Phys. Chem. Ref. Data,
 4, 493
\bibitem[]{} Fuhr, J. R., Wiese, W. L., 2005, CRC Handbook of Chemistry
 and Physics, 86$^{th}$ Edition
\bibitem[]{} Fuhr, J. R., Wiese, W. L., 2006, J. Phys. Chem. Ref. data,  35, 1669
\bibitem[]{} Gehren, T., Liang, Y. C., Shi, J. R., Zhang, H. W., Zhao, G., 2004, A\&A, 413, 1045
\bibitem[]{} Gielen, C., Van Winckel, H., Min, M., Waters, L. B. F. M., Lloyd Evans, T., 2008, A\&A, 490, 725
\bibitem[]{} Giridhar, S., Rao, N. K., Lambert, D. L., 1994, ApJ, 437, 476
\bibitem[]{} Giridhar, S., Arellano Ferro, A., Parrao, L., 1997, PASP, 109, 1077
\bibitem[]{} Giridhar, S., Lambert, D. L., Guillermo Gonzalez, 1998, 509, 366
\bibitem[]{} Giridhar, S., Lambert, D. L., Guillermo Gonzalez, 2000, ApJ, 531, 521
\bibitem[]{} Giridhar, S., Arellano Ferro, A., 2005, A\&A, 443, 297
\bibitem[]{} Giridhar, S., Lambert, D. L., Reddy, B. E., Guillermo Gonzalez., David Yong., 2005, ApJ, 627, 432
\bibitem[]{} Giridhar, S., Molina, R., Arellano Ferro, A., Selvakumar, G., 2010, MNRAS, 406, 290
\bibitem[]{} Guillermo Gonzalez, Lambert, D. L., Giridhar, S., 1997a, ApJ, 481, 452
\bibitem[]{} Guillermo Gonzalez, Lambert, D. L., Giridhar, S., 1997b, ApJ, 479, 427
\bibitem[]{} Hannaford, P., Lowe, R. M., Grevesse, N., Bi\'{e}mont, E., Whaling, W., 1982, ApJ, 261, 736
\bibitem[]{} Herwig, F., 2005, ARA\&A, 43, 435
\bibitem[]{} Hrivnak, J., 1995, ApJ, 438, 341
\bibitem[]{} Hrivnak, B. J., Van Winckel, H., Reyniers, M., Bohlender, D., Waelkens, C., Wenxian Lu., 2008, AJ, 136, 1557
\bibitem[]{} Kelleher, D. E., Podobedova, L. I., 2008a, J. Phy. Chem. Ref. Data,
 37, 267
\bibitem[]{} Kelleher, D. E., Podobedova, L. I., 2008b, J. Phy. Chem. Ref. Data,
 37, 709
\bibitem[]{} Kelleher, D. E., Podobedova, L. I., 2008c, J. Phy. Chem. Ref. Data,
 37, 1285
\bibitem[]{} Kiss, L. L., Derekas, A., Szab\'{o}, M., Bedding, T. R., Szabados, L., 2007, MNRAS, 375, 1338
\bibitem[]{} Klochkova, V. G., Panchuk, V. E., 1996, Bull.Spec.Astrophys.Obs,
 41, 5
\bibitem[]{} Klochkova, V. G., Panchuk, V. E., 1998, Astronomy Letters, 24, 650
\bibitem[]{} Klochkova, V. G., Panchuk, V. E., Tavolzhanskaya, N. S., 2010, Astronomy Reports, 54, 234
\bibitem[]{} Kodaira, K., Greenstein, J. L., Oke, J. B., 1970, ApJ, 159, 485.
\bibitem[]{} Korotin. S. A., 2009, Astronomy Reports, 53, 651 
\bibitem[]{} Kwok, S., 1993, ARA\&A, 31, 63
\bibitem[]{} Kurucz, R.L., Furenlid, I., Brault, J., Testerman,L.,1984,
 National Solar Observatory Atlas, Sunspot, New Mexico: National Solar Observatory
\bibitem[]{} Lawler, J. E., Bonvallet, G., Sneden, C., 2001a, ApJ,
556, 452
\bibitem[]{} Lawler, J. E., Wickliffe, M. E., Den Hartog, E. A., Sneden, C., 2001b, ApJ, 563, 1075
\bibitem[]{} Lawler, J. E., Den Hartog, E. A., Sneden, C., Cowan, J. J.,
 2006, ApJS, 162, 227
\bibitem[]{} Lawler, J. E., Sneden, C., Cowan, J. J., Ivans, I. I., 
Den Hartog, E. A., 2009, ApJS, 182, 51
\bibitem[]{} Lind, K., Asplund, M., Barklem, P. S., Belyaev, A. K., 2011, A \&A, 528, 103
\bibitem[]{} Luck, R. E., Lambert, D. L., Bond, H. E., 1983, PASP, 98, 413
\bibitem[]{} Luck, R.E., Bond, H.E., 1989, ApJ, 342, 476
\bibitem[]{} Lyubimkov, L. S., Lambert, D. L., Korotin, S. A., Poklad, D. B., Rachkovskaya, T. M., Rostopchin, S. I., 2011, MNRAS, 410, 1774
\bibitem[]{} Maas, T., Van Winckel, H., Waelkens, C., 2002, A\&A, 386, 504
\bibitem[]{} Maas, T., Van Winckel, H., Llyod Evans, T., Nyman, L. \AA~., Kilkenny, D., Martinez, P., Marang, F., van Wyk, F., 2003, A\&A, 405, 271 
\bibitem[]{} Maas, T., Van Winckel, H., Evans, L. T., 2005, A\&A, 429, 297
\bibitem[]{} Maas, T., Giridhar, S., Lambert, D. L., 2007, ApJ, 666, 378
\bibitem[]{} Martin, G. A., Fuhr, J. R., Wiese, W. L., 1988, J. Phys. Chem. Ref. Data, 3, 512
\bibitem[]{} Mashonkina, L., Korn, A. J., Przybilla, N., 2007, A\&A, 461, 261
\bibitem[]{} Mashonkina, L., 2011, arXiv:1104.4403 
\bibitem[]{} McWilliam, A., 1998, AJ, 115, 1640
\bibitem[]{} Mel\'{e}ndez, J., Barbuy, B., 2009, A\&A,
497, 611
\bibitem[]{} Mowlavi, N., 1999, A\&A, 350, 73
\bibitem[]{}  Mucciarelli, A., Caffau, E., Freytag, B., Hans-G\"{u}nter Ludwig, Bonifacio, P., 2008, A\&A, 484, 841
\bibitem[]{} Podobedova, L. I., Kelleher, D. E., Wiese, W. L., 2009,
J. Phys. Chem. Ref. Data, 38, 171
\bibitem[]{} Prochaska, J. X., McWilliam, A., 2000, ApJ, 537, 57
\bibitem[]{} Rao, N. K., Goswami, A., Lambert, D. L., 2002, BASI, 30, 671
\bibitem[]{} Rao, N. K., Reddy, B. E., 2005, MNRAS, 357, 235
\bibitem[]{} Reddy, B. E., Parthasarathy, M., 1996, AJ, 112, 2053
\bibitem[]{} Reddy, B. E., Bakker, E. J., Hrivnak, B. J., 1999, ApJ, 524, 831
\bibitem[]{} Reddy, B. E., Guillermo Gonzalez., David Yong, 2002, ApJ, 564, 482
\bibitem[]{} Reddy, B. E., Lambert, D. L., Allende Prieto, C., 2006, MNRAS, 367, 1329
\bibitem[]{} Reyniers, M., 2002, Ph.D Thesis, K. U. Leuven
\bibitem[]{} Reyniers, M., Van Winckel, H., Gallino, R., Straniero, O., 2004, A\&A, 417, 269
\bibitem[]{} Reyniers, M., Van de Steene, G. C., Van Hoof, P. A. M., Van Winckel, H., 2007, A\&A, 471, 247
\bibitem[]{} Romano, D., Karakas, A. I., Tosi, M., Matteucci, F., 2010, A\&A, 522, A32
\bibitem[]{}Sansonetti, J. E., 2008, J. Phys. Chem. Ref. Data, 37, 1659
\bibitem[]{} Schaller, G., Schaerer, D., Mainer, G., Maeder, A., 1992, A\&AS, 96, 269
\bibitem[]{} Schiller, F., Pryzbilla, N., 2008, A\&A, 479, 849
\bibitem[]{} Si\'{o}dmiak, N., Meixner, M., Ueta, T., Sugerman, B. E. K., Van de Steene, G. C., Szczerba, R., 2008, ApJ, 677, 382
\bibitem[]{} Sneden, C., 1973, Ph.D Thesis, Univ. of Texas at Austin, USA
\bibitem[]{} Sneden, C., Lawler, J. E., Cowan, J. J., Ivans, I. I., Den Hartog, E. A., 2009, ApJS, 182, 80 
\bibitem[]{} Sobeck, J. S., Lawler, J.E., Sneden, C., 2007, ApJ, 667,
 1267
\bibitem[]{} Su\'{a}rez, O., Garc\'{i}a-Lario, P., Manchado, A., Manteiga, M., Ulla, A., Pottasch, S. R., 
 2006, A\&A, 458, 173
\bibitem[]{} Szczerba, R., Si\'{o}dmiak, N., Stasi\'{n}ska, G., Borkowski, J.,   2007, A\&A, 469, 799
\bibitem[]{} Takeda, Y., Takada-Hidai, M., 1998, PASJ, 50, 629
\bibitem[]{} Takeda, Y., Sato, B., Murata, D., 2008, PASJ, 60, 781
\bibitem[]{} Th\'{e}venin, F., Idiart, T. P., 1999, ApJ, 521, 753
\bibitem[]{} Tull, R. G., MacQueen, P. J., Sneden, C., Lambert, D. L., 1995, PASP, 107, 251
\bibitem[]{} van der Veen, W. E. C. J., Habing, H. J., 1988, A\&A, 194, 125
\bibitem[]{} Van Winckel, H., 1995, Ph.D Thesis,  K. U. Leuven
\bibitem[]{} Van Winckel, H., 1996, A\&A, 319, 561
\bibitem[]{} Van Winckel, Waelkens, C., Waters, L. B. F. M., 1996, A\&A, 306, 37
\bibitem[]{} Van Winckel, H., 1997, A\&A, 319, 561
\bibitem[]{} Van Winckel, H., Reyniers, M., 2000, A\&A, 354, 135
\bibitem[]{} Van Winckel, H., Waelkens, C., Waters, L. B. F. M., 2000, IAUS, 177, 285
\bibitem[]{} Van Winckel, H., 2003, ARA\&A, 41, 391
\bibitem[]{} Velichko, A. B., Mashonkina, L. I., Nilsson, H., 2010, Astronomy Letters, 36, 664
\bibitem[]{} Venn, K.A., 1995, ApJ, 449, 839.
\bibitem[]{} Ventura, P., D'Antona, F., 2009, A\&A, 499, 835
\bibitem[]{} Wedemeyer, S., 2001, A\&A, 373, 998
\bibitem[]{} Wiese, W. L., Fuhr, J. R., Deters, T. M., 1996, J. Phy. Chem. Ref. Data, Monograph No. 7
\bibitem[]{} Wiese, W. L., Fuhr, J. R., 2007, J. Phy. Chem. Ref. Data,
 36, 1287

\end{thebibliography}
\end{document}